\documentclass[submission, Phys]{SciPost}

\hypersetup{
    colorlinks,
    linkcolor={red!50!black},
    citecolor={blue!50!black},
    urlcolor={blue!80!black}
}

\usepackage[bitstream-charter]{mathdesign}
\DeclareSymbolFont{usualmathcal}{OMS}{cmsy}{m}{n}
\DeclareSymbolFontAlphabet{\mathcal}{usualmathcal}

\newcommand{\Gm}{{\widehat{\varG}}}
\newcommand{\gm}{{\widehat{g}}}

\newcommand{\wPsi}{{\widetilde{\Psi}}}

\newcommand{\varA}{{\mathcal{A}}}
\newcommand{\varB}{{\mathcal{B}}}
\newcommand{\varG}{{\mathcal{G}}}
\newcommand{\varH}{{\mathcal{H}}}
\newcommand{\varI}{{\mathcal{I}}}
\newcommand{\varN}{{\mathcal{N}}}
\newcommand{\varO}{{\mathcal{O}}}
\newcommand{\varQ}{{\mathcal{Q}}}
\newcommand{\varT}{{\mathcal{T}}}

\newcommand{\bfk}{{\mathbf{k}}}
\newcommand{\bfG}{{\mathbf{G}}}
\newcommand{\bfGamma}{{\mathbf{\Gamma}}}
\newcommand{\bfSigma}{{\mathbf{\Sigma}}}
\newcommand{\bfXi}{{\mathbf{\Xi}}}

\newcommand{\avg}[1]{\left\langle#1\right\rangle}
\newcommand{\Avg}[1]{\langle#1\rangle}
\newcommand{\comm}[2]{[#1,#2]}
\newcommand{\acomm}[2]{\{#1,#2\}}

\newcommand{\rcp}[1]{\frac{1}{#1}}

\newcommand{\ode}[3][]{\frac{d^{#1}{#2}}{d{#3}^{#1}}}
\newcommand{\Ode}[2][]{\frac{d^{#1}}{d{#2}^{#1}}}

\makeatletter
\newcommand{\Tr}{\mathop{\operator@font Tr}}
\newcommand{\im}{\mathop{\operator@font Im}}
\newcommand{\re}{\mathop{\operator@font Re}}
\makeatother

\let\up\uparrow
\let\down\downarrow

\newcommand{\eqnref}[1]{Eq.~(\ref{#1})}
\newcommand{\eqnsref}[1]{Eqs.~(\ref{#1})}

\newcommand{\Eqnsref}[1]{Equations~(\ref{#1})}

\newcommand{\figref}[1]{Fig.~\ref{#1}}
\newcommand{\figsref}[1]{Figs.~\ref{#1}}
\newcommand{\Figref}[1]{Figure~\ref{#1}}

\newcommand{\secref}[1]{Sec.~\ref{#1}}
\newcommand{\secsref}[1]{Secs.~\ref{#1}}

\newcommand{\Secsref}[1]{Sections~\ref{#1}}

\newcommand{\appref}[1]{Appendix~\ref{#1}}

\newcommand{\Left}{l}
\newcommand{\Right}{r}

\begin{document}

\begin{center}
  {\Large \textbf{
      Heat and charge transport in interacting nanoconductors driven by
      time-modulated temperatures
    }}
\end{center}

\begin{center}
Rosa L\'{o}pez\textsuperscript{1},
Pascal Simon\textsuperscript{2} and
Minchul Lee\textsuperscript{3$\star$}
\end{center}

\begin{center}
  {\bf 1} Institut de F\'{\i}sica Interdisciplin\`aria i de Sistemes Complexos
  IFISC (CSIC-UIB), E-07122 Palma de Mallorca, Spain
  \\
  {\bf 2} Universit\'e Paris-Saclay, CNRS, Laboratoire de Physique des Solides,
  91405, Orsay, France
  \\
  {\bf 3} Department of Applied Physics and Institute of Natural Science,
  College of Applied Science, Kyung Hee University, Yongin 17104, Korea
  \\
  ${}^\star$ {\small \sf minchul.lee@khu.ac.kr}
\end{center}

\begin{center}
\today
\end{center}


\section*{Abstract}
{\bf


We investigate the quantum transport of the heat and the charge through a
quantum dot coupled to fermionic contacts under the influence of time modulation of temperatures.
We derive, within the nonequilibrium Keldysh Green's function formalism,
generic formulas for the charge and heat currents by extending the concept of
gravitational field introduced by Luttinger to the dynamically driven system
and by identifying the correct form of dynamical contact energy.
In linear response regime our formalism is validated from satisfying the
Onsager reciprocity relations and demonstrates its utility to reveal nontrivial
dynamical effects of the Coulomb interaction on charge and energy relaxations.
}

\vspace{10pt}
\noindent\rule{\textwidth}{1pt}
\tableofcontents\thispagestyle{fancy}
\noindent\rule{\textwidth}{1pt}
\vspace{10pt}

\section{Introduction}
Nanoelectronics \cite{Datta2005} and the emerging field of Thermotronics
\cite{Wang2007,Dhar2008,Li2012,BenAbdallah2017} are at the forefront of
manipulating electron charge and energy fluxes through electrostatic and
thermal gradients, respectively. The interplay between these technologies,
known as Thermoelectricity \cite{Benenti2017} explores how a thermal gradient
influences charge current and vice versa, expanding the functionality of
quantum conductors. Investigating time-dependent transport in nanostructures
unlocks unique insights not achievable with static fields
\cite{Buttiker1993jun,Jauho1994aug,Platero2004}. The ability to control these
nanoscale systems with electrical or thermal drivings paves the way for new
possibilities in quantum technologies. Electrically modulated quantum
conductors lead to innovative devices like electron pumps
\cite{Kouwenhoven1991sep,Howe2021,Blumenthal2007,Giblin2012,Rossi2014,Tettamanzi2014},
dynamical Coulomb Blockade quantum systems \cite{Grabert1992,Duprez2021may},
and AC-driven nano-electromechanical systems for sensors \cite{Blick2005}. This
setup can function as a quantized emitter, operating as a quantum capacitor in
the adiabatic regime and behaving like an RC circuit in the GHz range, with a
unique charge relaxation resistance quantized as $R_0 = h/2e^2$
\cite{Buttiker1993jun,Gabelli2006,Parmentier2012apr}. They have demonstrated to
serve as single electron sources for quantum computing applications and
metrology
\cite{Pekola2013,Averin2008,McNeil2007,Blumenthal2007,Kaestner2008,Feve2007,Brunpicard2016}.

Similarly, thermally modulated nanoconductors, within the framework of quantum thermodynamics  \cite{Gemmer2009}, are rapidly advancing. Examples include quantum thermal machines \cite{Holubec2020,Martinez2015,Scopa2019,Martinez2016,Rossnagel2016,Brunpicard2016}, thermal diodes  \cite{Li2004,Otey2010}, thermal transistors \cite{Li2006,BenAbdallah2013,BenAbdallah2014,Joulain2015}, thermal memristors \cite{DiVentra2009} and thermal capacitors \cite{Li2012,Dhar2008,Wang2007}. Despite significant progress, time-dependent transport in nanostructures, especially in thermally driven systems, remains a challenging and vibrant field. Thus far, researchers have primarily tackled this issue in two scenarios: the incoherent transport regime \cite{Portugal2021nov} and the adiabatic driving \cite{Thouless1983may,Buttiker1994,Pretre1996sep,Brouwer1998oct}, characterized by a slow time-modulation leading to a net transport of charge or heat.
Adiabatic pumping for the design of thermal machines \cite{Brandner2020jan,Bhandari2020oct} is rooted in geometrical concepts like the Berry connection and it has very broad applications in electronics, thermotronics \cite{BenAbdallah2017}, and quantum information processing. Recent studies on temperature-driven dynamics in two-level systems engineer an effective temperature through oscillator frequency manipulation \cite{Portugal2022nov}.
However, most of the existing temperature-driven studies heavily rely on scattering theory, lacking the consideration of Coulomb interaction beyond mean-field treatment and being unable to capture dynamical excitations induced by nonadiabatic temperature driving in the presence of nontrivial interactions.

New and exciting functionalities are expected to arise in time-dependent
thermally-driven interacting nanoconductors away from the adiabatic
regime. This issue is the central objective of our work. The primary challenge
in introducing a thermal bias lies in achieving a microscopic formulation. In
this context, reconciling the macroscopic nature of temperature gradients and
thermal forces, which arise from statistical averaging, appears to be
incompatible. Then, representing these effects in a microscopic quantum
mechanical Hamiltonian is not straightforward. However, in 1964, Luttinger
\cite{Luttinger1964,Hasegawa2018} presented a clever and ingenious solution. He
introduced a scalar potential, $\Psi$, referred as a \textit{gravitational
  field}, which couples to the energy density of the system, $J_E$. This scalar
field can be viewed as a mechanical conjugate to the energy density of the system
defined as follows
\begin{equation}
  \varH_G = \int J_E(\boldsymbol{r})\Psi(\boldsymbol{r}) d\boldsymbol{r}.
\end{equation}
Luttinger justified his trick by the request for this scalar field to satisfy the Einstein relation, i.e., the potential adjusts itself to balance the thermal force, resulting in an identity $\nabla\Psi =\nabla T/T$ in the thermal equilibrium. In this respect, the gravitational field or thermomechanical potential serves as a local proxy for local temperature variations.
This theoretical treatment finds application in determining linear responses to static gradients by introducing the scalar field \cite{Eich2014,Lozej2018aug}  or equivalently, a vector potential \cite{Tatara2015may}. It proves effective in various scenarios, including classical systems \cite{Shastry2008}, thermoelectrical transport in
quantum systems using density-functional theory
framework \cite{Eich2014} and transient current calculations  \cite{Eich2016,Lozej2018aug} in the stationary regime. In the latter, the adiabatic limit for time-dependent temperature is analyzed using the nonequilibrium Keldysh Green function (NEGF) formalism \cite{Hasegawa2017,Hasegawa2018,Brandner2020jan,Bhandari2020oct}, employing Luttinger's trick.

\begin{figure}[t]
  \centering
  \includegraphics[width=0.45\textwidth]{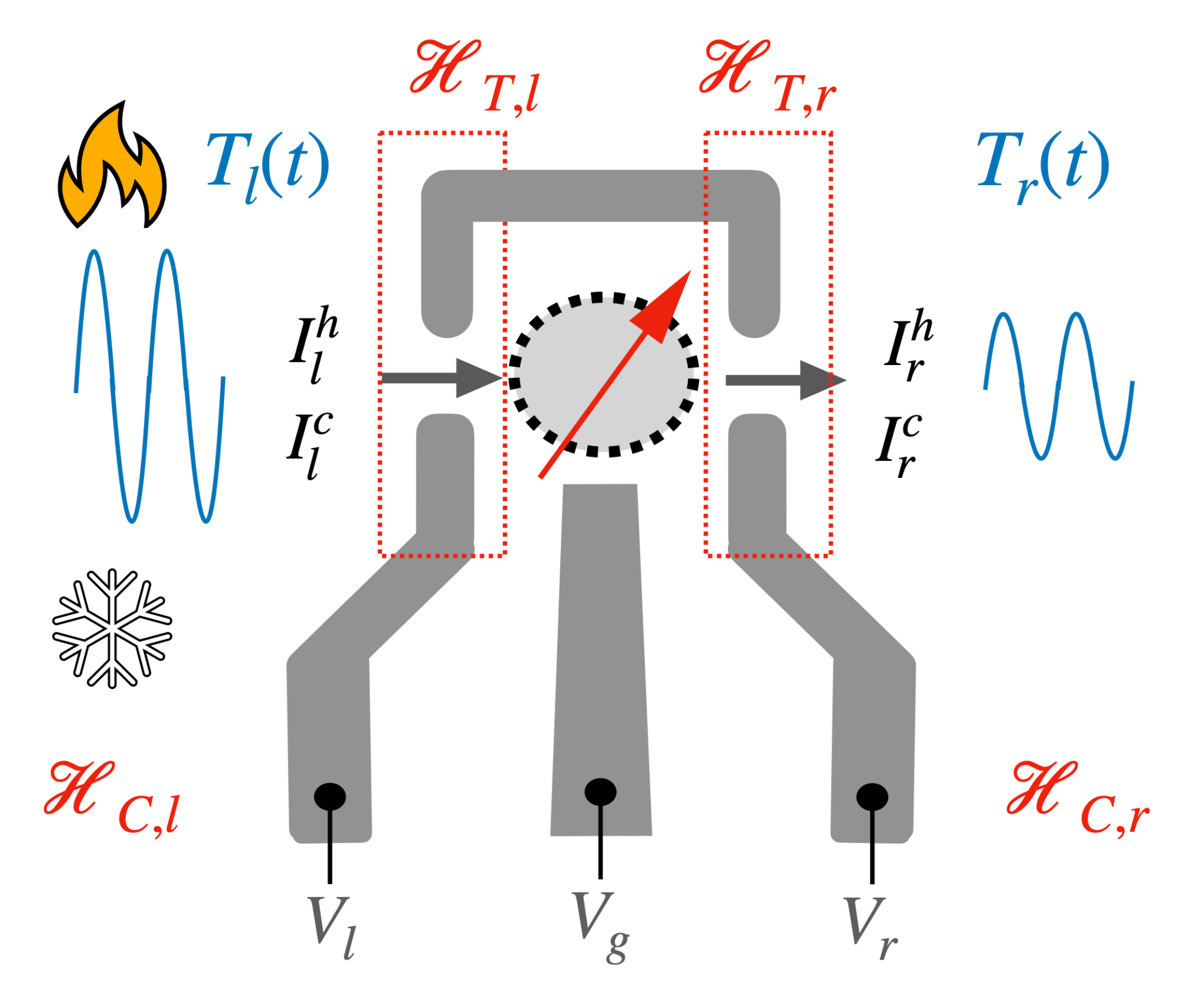}
  \caption{Lateral quantum dot system coupled to a left reservoir ($\Left$) and
    a right reservoir ($\Right$) that are described by $\varH_{{\rm C},\Left}$
    and $\varH_{{\rm C},\Right}$. Each reservoir is under the influence of a
    modulated temperature in time with $T_{\Left}(t)$ and $T_{\Right}(t)$. Left
    and right tunneling barriers are described by $\varH_{{\rm T},\Left}$ and
    $\varH_{{\rm T},\Right}$, respectively, as indicated. The central part
    corresponds a spinful quantum dot. Plunger gates $V_{\Left}$ and
    $V_{\Right}$ control the barrier transparency, an additional gate $V_g$ is
    applied to the quantum dot region to tune the dot level position denoted by
    $\epsilon_\sigma$.}
  \label{fig:scheme}
\end{figure}

Our work applies Luttinger's idea to a quantum conductor tunnel-coupled to two
electronic reservoirs subjected to time-dependent temperature modulation [see
\figref{fig:scheme}]. We focus on a quantum dot (QD), the most generic quantum
conductor, featuring multiple local levels with diverse interactions such as
Coulomb interaction and spin-orbit coupling. Our goal is to derive generic
expressions of the electric and heat currents flowing through the QD junctions
in the framework of the NEGF formalism. For that purpose we adopt the
gravitational field description to address the temperature modulation in time.
Our formalism to be proposed is quite widely applicable in that it can consider
diverse and complex quantum conductors and, in particular, it allows one to
deal with the effect of strong Coulomb interaction in a systematic way, in
contrast to previous methods. Also, it can be easily extended to involve
geometric configurations such as quantum mechanical interferometers.


Our formalism, in the course of its development, addresses a very important
technical aspect regarding the correct definition of the dynamical contact
energy and its coupling to the gravitational field. We adopt a tight-binding
model to describe the system, which divides the total Hamiltonian into pieces,
namely, the contact (reservoir) regions, the quantum dot, and the tunneling
barriers connecting the contacts to the QD. The contact energy responsible for
the heat current in the corresponding contact
\begin{equation}
  \label{eq:Q:0}
  \varQ_\ell
  = \varH_{\rm C\ell}+ \lambda \varH_{\rm T\ell}
  \quad\text{with }\ \lambda = 1/2
\end{equation}
should then include not only the energy stored in the contact
$\ell = \Left,\Right$ (described by the Hamiltonian $\varH_{\rm C\ell}$, see
\eqnref{eq:HC}) but also the half of the energy stored in the tunneling barrier
coupled to that contact (described by the Hamiltonian $\varH_{\rm T\ell}$, see
\eqnref{eq:HT}) \cite{Ludovico2014apr,Ludovico2016}.
Together with additional physical requirements (to be explained in
\secref{sec:NEGF}) it is natural to couple the Luttinger field to the contact
energy $\varQ_\ell$ (which therefore includes $\varH_{\rm T\ell}$).  This is
the most important ingredient for a correct application of the Luttinger's
trick to the calculation of the dynamical heat current.

Accommodating the Luttinger's trick and employing the NEGF technique, we
formulate expressions for charge and heat currents with a particular emphasis
on the linear response regime. Notably, our formalism yields comprehensive
expressions for the currents in relation to the QD NEGFs [see
\eqnsref{eq:Ic:linear} and (\ref{eq:Ih:linear:final})]. In special conditions,
by a help of the charge conservation and a sum rule on the energy change rates,
the currents can be obtained solely in terms of the retarded and advanced
components of the QD NEGFs [see \eqnref{eq:Ic:linear:final}], which streamlines
the computation of the currents.
It should be noted that our expressions for currents are the analogue to those
for electrical current induced by a time-dependent electrical modulation
originally presented in the pioneering work by Jauho, Wingreen and Meir
\cite{Jauho1994aug}, but in our case the time-dependent modulation is done
thermally instead.

In demonstrating our Luttinger formalism, we initially apply it to the
noninteracting case in which the fulfillment of the Onsager reciprocity
relation validate our approach and moreover our results align with those
obtained from scattering theory. Extending our formalism to the interacting
case within the Hartree approximation reveals that the Coulomb interaction can
alter the response for charging and energy relaxations with distinct
temperature dependences.  The success of our formalism in these applications
suggests its potential for studying dynamical heat transport in the presence of
the Coulomb blockade effect or many-body correlations.

Our paper is organized as follows: In \secref{sec:model} we introduce our model
Hamiltonian and implement the Luttinger's trick onto it. We also define the
charge and heat currents and find out the relevant sum rules. In
\secref{sec:NEGF} we express the charge and heat currents in terms of QD NEGFs
by solving the corresponding Dyson's equations for the NEGF.  Later, we
restrict ourselves to the linear response regime and further simplify the
expressions for the charge and heat currents with a help of sum rules, which
are the main results of our work. \Secsref{sec:ni} and \ref{sec:hartree}
demonstrate the applications of our Luttinger formalism, considering the
noninteracting and interacting cases, respectively. In \secref{sec:conclusion},
we summarize our work and discuss the possible applications and extensions of
our formalism and the experimental proposals.

\section{Thermal Fields and Currents}
\label{sec:model}

For our study, we consider a quantum dot coupled to two fermionic
reservoirs. To represent this configuration, we establish the Hamiltonian in
the context of a tight-binding model. To construct this Hamiltonian, we first
consider the distinct components that comprise our nanoconductor, and then
provide an explanation for how to add the term corresponding to the
gravitational field. The tight-binding Hamiltonian comprises three
contributions. First, the Hamiltonian for the electrodes, denoted as $\ell$
that takes the values of left ($\Left$) and right ($\Right$), is given by
\begin{align}
  \label{eq:HC}
  \varH_{\rm C}
  = \sum_\ell \varH_{\rm C\ell}
  =
  \sum_\ell \sum_{\bfk\sigma} \epsilon_{\ell\bfk}
  c_{\ell\bfk\sigma}^\dag c_{\ell\bfk\sigma}\,,
\end{align}
where $c_{\ell\bfk\sigma}^\dag$ is the creation operator for an electron in the
lead $\ell$ with wavevector $\bfk$, spin $\sigma$($= \up,\down$), and energy
$\epsilon_{\ell\bfk}$ measured with respect to the Fermi level.
Second, the central conductor considered as an interacting localized
multi-level quantum dot is described as
\begin{align}
  \label{eq:HD}
  \varH_{\rm D}
  = \varH_{\rm D}(\{d_m,d_m^\dag\})
\end{align}
Here $d_m$ annihilates an electron on the localized level $m$
$(m=1,\cdots,N_d)$ where the index $m$ denotes both the orbital level and the
spin.  $\varH_{\rm D}$ can describe any kind of interactions inside the quantum
dot such as the Coulomb interactions and the spin-orbit couplings.
Thirdly, the tunneling Hamiltonian that connects each of the two electrodes
with the central conductor reads
\begin{align}
  \label{eq:HT}
  \varH_{\rm T}
  = \sum_\ell \varH_{\rm T\ell}
  = \sum_\ell \sum_{m\bfk\sigma}
  \left(
    t_{\ell\bfk\sigma,m} d_m^\dag c_{\ell\bfk\sigma}
    +
    t_{\ell\bfk\sigma,m}^* c_{\ell\bfk\sigma}^\dag d_m
  \right),
\end{align}
where the tunneling amplitude between the lead $\ell$ and the dot level $m$ is
denoted by $t_{\ell\bfk\sigma,m}$.  Hereafter, for simplicity, we assume that
the tunneling amplitude is momentum-independent
$t_{\ell\bfk\sigma,m} \equiv t_{\ell\sigma,m}$. Accordingly, we define the
escaping tunneling rates
$\Gamma_{\ell,mm'} \equiv \sum_\sigma \pi\rho_0t_{\ell\sigma,m}
t_{\ell\sigma,m'}^*/\hbar$ for level $m$ and $m'$ due to the coupling to the
contact $\ell$, where $\rho_0$ is the density of states in the contacts at the
Fermi energy. Later it is convenient to express the rates in the form of a
$N_d\times N_d$ matrix $\bfGamma_\ell$ whose matrix elements are given by $\Gamma_{\ell,mm'}$.

\subsection{Luttinger's trick and Hamiltonian}

The idea about the Luttinger's trick consists of introducing new fields, dubbed
as \emph{gravitational fields} $\Psi_\ell(t)$ which are coupled to the contact
energies. We determine the precise form of this coupling based on the following \textit{two}
arguments. First, as mentioned before, the contact energies in the framework of
a tight-binding formulation should be redefined to account for the reactance
or energy stored in the barrier \cite{Ludovico2014apr}, leading to
$\varQ_\ell$ as specified in \eqnref{eq:Q:0}. Therefore, in our setup, the term
responsible for the dynamical thermal driving is introduced as
\begin{align}
  \label{eq:HG}
  \varH_{\rm G}
  = \sum_\ell \Psi_\ell(t) \left(\varQ_\ell - \avg{\varQ_\ell}_0\right),
\end{align}
where each of the $\Psi_\ell(t)$ field is coupled to the \textit{excess} energy for the
contact $\ell$ with respect to its equilibrium value (at $\Psi_\ell = 0$);
$\avg{\cdots}_0$ denotes the expectation value at equilibrium. The coupling to
the excess energy is reasonable because for sufficiently weak driving or at low
temperature only the excitations around the Fermi level will contribute to the
electric and thermal transports. Furthermore, mathematically, it introduces
only the additional time-varying number,
$-\sum_\ell \Psi_\ell(t) \avg{\varQ_\ell}_0$ which does not affect the dynamics
of states.
The constant $\lambda$ in $\varQ_\ell$, the coupling coefficient of the
gravitational field to the tunneling barrier energy, is set to be $1/2$, but
for a time being we keep this symbol as it is in order to track down how the precise value of this
coefficient influences the results.

Second, we require that the coupling to the gravitational field should not
affect the effective coupling between the dot and the leads
\cite{Hasegawa2017,Hasegawa2018} since the dot-lead coupling should be immune
to the temperature in the leads. Interestingly, we find that the coupling in Eq. (\ref{eq:HG}) automatically satisfies this requirement, at least, in the linear
response regime, as we will see later [see \secref{sec:linear_response_regime}].

The complete Hamiltonian for our setup is then described by
\begin{align}
  \label{eq:H}
  \varH(t)
  = \varH_{\rm C} + \varH_{\rm D} + \varH_{\rm T} + \varH_{\rm G}(t)
  = \varH_{\rm C,\Psi}(t) + \varH_{\rm D} + \varH_{\rm T,\Psi}(t),
\end{align}
where we have introduced the time-dependent contact and tunneling Hamiltonians
defined as
\begin{align}
  \varH_{\alpha,\Psi}(t)
  \equiv \sum_\ell \varH_{\alpha\ell,\Psi}(t)
  =
  \sum_\ell
  \left[
    \varH_{\alpha\ell}
    + \lambda_\alpha \Psi_\ell(t) (\varH_{\alpha\ell} - \avg{\varH_{\alpha\ell}}_0)
  \right],
\end{align}
for $\alpha = \rm C,T$ and with $\lambda_{\rm C} = 1$ and
$\lambda_{\rm T} = \lambda$.  We assume that the thermal drivings on both the
contacts are periodic with the same frequency $\Omega = 2\pi/\tau$:
$\Psi_\ell(t+\tau) = \Psi_\ell(t)$. More specifically, we take a sinusoidal
time dependence:
\begin{align}
  \label{eq:Psi}
  \Psi_\ell(t) = \Psi_\ell \cos \Omega t.
\end{align}
Here the factors $\Psi_\ell$ are real and can be zero.
Since the dynamical variation of the temperature is taken into account via the
gravitaional field, both the contacts are assumed to have the same chemical
potential and the same base temperature $T$ (or the inverse temperature
$\beta$) so that the thermal populations of both the uncoupled contacts are
specified by the same Fermi distribution function $f(\epsilon)$.

\subsection{Charge currents and  charge conservation}

We focus in our study on the charge and heat currents transversing the
contacts, driven by the dynamical change of the temperature. The charge
currents, or the change rates of the charges in the contacts
($\ell=\Left,\Right$) are described in terms of the charge current operators
\begin{equation}
  \varI^c_\ell
  \equiv -e \ode{\varN_\ell}{t}
  = -\frac{ie}{\hbar} \comm{\varH}{\varN_\ell}
\end{equation}
where
$\varN_\ell \equiv \sum_{\bfk\sigma} c_{\ell\bfk\sigma}^\dag
c_{\ell\bfk\sigma}$ are the charge number operators for the contact
$\ell$. Under this consideration, the charge currents, the expectation values
of the charge current operators for the contacts and the quantum dot can be
obtained by evaluating
\begin{subequations}
  \label{eq:I:0}
  \begin{align}
    I^c_\ell(t)
    & =
    - \frac{ie}{\hbar}
    \sum_{m\bfk\sigma} t_{\ell\bfk\sigma,m}(t) \Avg{d_m^\dag(t) c_{\ell\bfk\sigma}(t)}
    + (c.c),
    \\
    I^c_{\rm D}(t)
    & = - e \ode{\avg{\varN_{\rm D}}}{t}
    = - e \sum_m \ode{\avg{n_m}}{t},
  \end{align}
\end{subequations}
where $\varN_{\rm D} \equiv \sum_m n_m$ (with $n_m \equiv d^\dag_m d_m$) is the
charge number operator for the quantum dot.
%
It should be noted that the charge conservation condition,
$\comm{\varH}{\sum_\ell \varN_\ell + \varN_{\rm D}} = 0$, guarantees that
their sum vanishes at all time $t$:
\begin{align}
  \label{eq:chargeconservation}
  \sum_\ell I^c_\ell(t) + I^c_{\rm D}(t) = 0.
\end{align}
Later we will use this equality to simplify our final results.

\subsection{Heat currents and sum rule}

The heat currents for the electrodes are derived from the time derivative of
the energies stored at the contacts. Then, according to our choice of the
contact energy $\varQ_\ell$ which incorporates the contribution from the
neighboring tunneling barrier, the heat current for contact $\ell$ is given by
\cite{Ludovico2014apr,Ludovico2016}
\begin{align}
  \label{eq:Ih:0}
  I^h_\ell(t)
  = \ode{\avg{\varQ_\ell(t)}}{t}.
\end{align}
It should be noted \cite{note1} that multiple choices for the definition of the
heat currents for contacts are {\it a priori} possible in the Luttinger
formalism, while we find that \eqnref{eq:Ih:0} is the most suitable one.

The power supplied by the time-dependent thermal source or the power dissipated
is defined as
$P(t)=\Avg{\partial H/\partial t}=\Avg{\partial H_{\rm G}/\partial t}$ and
explicitly expressed as
\begin{align}\label{eq:P:0}
  P(t)
  =
  \sum_\ell \dot\Psi_\ell \left(\avg{\varQ_\ell} - \avg{\varQ_\ell}_0\right)
\end{align}
which contains the source contributions from the contacts and the tunneling
barriers.

Under the thermal driving, the energy conservation does not hold: $P(t) \ne
0$. However, one can still find an equality similar to
\eqnref{eq:chargeconservation}. To this purpose, we define the energy change
rates as
\begin{subequations}
  \begin{align}
    W_{\alpha\ell}(t)
    & \equiv
    \frac{i}{\hbar} \comm{\varH(t)}{\varH_{\alpha\ell,\Psi}(t)}
    \quad(\alpha = \rm C,T),
    \\
    W_{\rm D}(t)
    & \equiv
    \frac{i}{\hbar} \comm{\varH(t)}{\varH_{\rm D}}.
  \end{align}
\end{subequations}
Then, from \eqnref{eq:H}, the obvious commutation relation
$\comm{\varH(t)}{\varH(t)} = 0$ leads to a useful equality
\begin{align}
  \label{eq:Wsum}
  \sum_\ell (W_{\rm C\ell}(t) + W_{\rm T\ell}(t)) + W_{\rm D}(t) = 0
\end{align}
which is to be exploited later.

\section{Charge and Heat Currents in terms of NEGF´s}
\label{sec:NEGF}


In the seminal research \cite{Jauho1994aug}, the charge current flowing through
a quantum dot under a dynamical electric drive  was formulated in
a closed form in terms of the interacting QD Green functions. In a similar way,
 we formulate here the charge and heat currents in terms of solely the QD NEGFs
when the system is driven by oscillating temperatures. For such purpose we
adopt the nonequilibrium Keldysh formalism and employ the equation-of-motion
technique together with the Langreth rules \cite{Langreth1976}. Throughout our
paper, we use the retarded/advanced and lesser QD NEGFs defined as
\begin{subequations}
  \label{eq:GF:QD}
  \begin{align}
    \varG^{R/A}_{mm'}(t,t')
    & \equiv
    \mp i \Theta(\pm(t-t')) \Avg{\acomm{d_m(t)}{d_{m'}^\dag(t')}},
    \\
    \varG^<_{mm'}(t,t')
    & \equiv i \Avg{d_{m'}^\dag(t') d_m(t)},
  \end{align}
\end{subequations}
respectively, and the contact and QD-contact NEGFs are defined in a similar
way: For examples,
\begin{subequations}
  \label{eq:GF:contact-QD}
  \begin{align}
    \varG^<_{\ell\bfk\sigma,\ell'\bfk'\sigma'}(t,t')
    & \equiv i\Avg{c_{\ell'\bfk'\sigma'}^\dag(t') c_{\ell\bfk\sigma}(t)},
    \\
    \varG^<_{m,\ell\bfk\sigma}(t,t')
    & \equiv i\Avg{c_{\ell\bfk\sigma}^\dag(t') d_m(t)}.
  \end{align}
\end{subequations}


Based on the definition of the Green's functions, the charge currents
(\ref{eq:I:0}) and the expectation values for the contact energy operators
(\ref{eq:Q:0}) are readily expressed in terms of the NEGF´s:
\begin{subequations}
  \begin{align}
    \label{eq:Ic:1}
    I^c_\ell(t)
    & =
    e
    \sum_{m\bfk\sigma}
    \frac{t_{\ell\bfk\sigma,m}^*(t)}{\hbar} \varG^<_{m,\ell\bfk\sigma}(t,t)
    + (c.c.),
    \\
    I^c_{\rm D}(t)
    & = e \sum_\sigma i \Ode{t} \varG^<_{mm}(t,t),
  \end{align}
\end{subequations}
and
\begin{subequations}
  \begin{align}
    \label{eq:HC:1}
    \avg{\varH_{\rm C\ell}}
    & =
    -i \sum_{\bfk\sigma} \epsilon_{\ell\bfk} \varG^<_{\ell\bfk\sigma,\ell\bfk\sigma}(t,t),
    \\
    \label{eq:HT:1}
    \avg{\varH_{\rm T\ell}}
    & =
    - i
    \sum_{m\bfk\sigma}
    t_{\ell\bfk\sigma,m}^* \varG^<_{m,\ell\bfk\sigma}(t,t)
    + (c.c.).
  \end{align}
\end{subequations}
Note that above we have introduced a time-varying effective tunneling
amplitudes,
\begin{align}
  \label{eq:t}
  t_{\ell\bfk\sigma,m}(t) \equiv (1 + \lambda \Psi_\ell(t)) t_{\ell\bfk\sigma,m}
\end{align}
for convenience. Knowing the time dependence of two expectation values,
$\avg{\varH_{\rm C\ell}}$ and $\avg{\varH_{\rm T\ell}}$, one can evaluate not
only the heat currents (\ref{eq:Ih:0}) and the power (\ref{eq:P:0}) but also
the energy change rates for contact $\ell$ and the tunneling parts via
\begin{equation}
  \label{eq:W:2}
  W_{\alpha\ell}(t)
  = (1 + \lambda_\alpha\Psi_\ell(t)) \ode{\avg{\varH_{\rm \alpha\ell}}}{t}
  \quad(\alpha = \rm C,T).
\end{equation}

Now, by employing the equation-of-motion technique \cite{Jauho1994aug}, the
lesser Green's functions, $\varG^<_{m,\ell\bfk\sigma}(t,t)$ and
$\varG^<_{\ell\bfk\sigma,\ell\bfk\sigma}(t,t)$, can be cast in terms of solely
the QD NEGFs [see \appref{app:FG} for details]. After some algebraic
manipulations on $\varG^<_{m,\ell\bfk\sigma}(t,t)$, the contact charge current
(\ref{eq:Ic:1}) and the expectation value of energy stored in the tunneling
barrier (\ref{eq:HT:1}) have compact forms in terms of self energies and the QD
NEGFs:
\begin{subequations}
  \label{eq:IcHT}
  \begin{align}
    \label{eq:Ic:2}
    I^c_\ell(t)
    & =
    e
    \int dt'
    \Tr
    \left[
      \bfG^R(t,t') \bfSigma^<_\ell(t',t)
      +
      \bfG^<(t,t') \bfSigma^A_\ell(t',t)
    \right]
    + (c.c.)
    \\
    \label{eq:HT:2}
    \avg{\varH_{\rm T\ell}}
    & =
    -
    \frac{i\hbar}{1+\lambda\Psi_\ell(t)}
    \int dt'
    \Tr
    \left[
      \bfG^R(t,t') \bfSigma^<_\ell(t',t)
      +
      \bfG^<(t,t') \bfSigma^A_\ell(t',t)
    \right]
    + (c.c.).
  \end{align}
\end{subequations}
Here $\bfG^a$ and $\bfSigma^a$ ($a=R,A,<$) are the $N_d\times N_d$ matrix
representations of the QD Green's functions and the self energies,
respectively. The traces are done over the QD orbital degrees of freedoms.
The matrix elements of the self energies are defined as
\begin{align}
  \label{eq:Sigma}
  \Sigma^a_{\ell,mm'}(t,t')
  \equiv
  \sum_{\bfk\sigma}
  \frac{t_{\ell\bfk\sigma,m}(t)}{\hbar}
  g^a_{\ell\bfk\sigma}(t,t')
  \frac{t_{\ell\bfk\sigma,m'}^*(t')}{\hbar},
\end{align}
where $g^a_{\ell\bfk\sigma}(t,t')$ with $a=R,A,<$ correspond to the
\textit{uncoupled} contact Green's functions governed by the time-dependent
Hamiltonian,
$\epsilon_{\ell\bfk}(t) c_{\ell\bfk\sigma}^\dag c_{\ell\bfk\sigma}$, with the
time-varying energy,
\begin{align}
  \label{eq:epsilon}
  \epsilon_{\ell\bfk}(t)
  \equiv (1 + \Psi_\ell(t)) \epsilon_{\ell\bfk}.
\end{align}
It should be noted that the gravitational field $\Psi_\ell(t)$ enters into the
self energy through the time-varying tunneling amplitude $t_{\ell\bfk}(t)$ as
well as the dynamically driven contact energy $\epsilon_{\ell\bfk}(t)$.

Similarly, by expressing $\varG^<_{\ell\bfk\sigma,\ell\bfk\sigma}(t,t)$ in
terms of the QD NEGFs, the expectation value of the contact Hamiltonian is
written as
\begin{align}
  \label{eq:HC:2}
  \begin{split}
    \avg{\varH_{C\ell}}
    -
    E_{\rm C\ell0}
    & =
    \int dt' \int dt''
    \Tr
    \left[
      \bfXi^{AR}_\ell(t,t'',t')
      \bfG^<(t',t'')
    \right.
    \\
    & \qquad\qquad\qquad\qquad\quad\left.\mbox{}
      +
      \bfXi^{<R}_\ell(t,t'',t')
      \bfG^R(t',t'')
      +
      \bfXi^{A<}_\ell(t,t'',t')
      \bfG^A(t',t'')
    \right],
  \end{split}
\end{align}
where
$E_{\rm C\ell0} \equiv -i \sum_{\bfk\sigma} \epsilon_{\ell\bfk}
g^<_{\ell\bfk\sigma}(t,t)$ are the \textit{equilibrium} energies contained in
the \textit{unperturbed} contacts. Note that this constant is irrelevant in our
study of heat flow because it does not contribute to the heat current once the
time derivative is taken. The matrix elements of the self-energy-like term
$\bfXi^{ab}_\ell$ in \eqnref{eq:HC:2} are defined as
\begin{align}
  \label{eq:Xi}
  \Xi^{ab}_{\ell,m'm''} (t,t',t'')
  \equiv
  -i \sum_{\bfk\sigma}
  \frac{t_{\ell\bfk\sigma,m'}(t')}{\hbar}
  g^a_{\ell\bfk\sigma}(t',t)
  \epsilon_{\ell\bfk}
  g^b_{\ell\bfk\sigma}(t,t'')
  \frac{t_{\ell\bfk\sigma,m''}^*(t'')}{\hbar}
\end{align}
with $a,b = R, A, <$.

In the presence of the time-dependent terms in the Hamiltonian, the Green's
functions $\varG(t,t')$ depend not on the time difference $t-t'$ but on $t$ and
$t'$ separately. However, since the driving is periodic with period $\tau$, the
Green's functions and the self energies are also periodic with
$\varG(t+\tau,t'+\tau) = \varG(t,t')$ so that it is more convenient to apply
the Fourier transformation and to move from the time domain to the frequency
domain. We adopt the mixed time-energy representation for the Fourier
transformation:
\begin{align}
  \label{eq:FT}
  \varG(t,t')
  =
  \int_{-\infty}^\infty \frac{d\omega}{2\pi} e^{-i\omega(t-t')} \varG(t,\omega)
  \quad\text{and}\quad
  \varG(t,\omega)
  = \sum_{n=-\infty}^\infty \varG(n,\omega) e^{-in\Omega t}.
\end{align}
Then, the integrals in \eqnsref{eq:IcHT} and (\ref{eq:HC:2}) can be expressed
in terms of the Fourier components:
\begin{subequations}
  \begin{align}
    \label{eq:FTapp:1}
    \int dt' \Tr\left[\bfG^a(t,t') \bfSigma^b_\ell(t',t)\right]
    & =
    \sum_{nn'} \int \frac{d\omega}{2\pi} e^{-i(n+n')\Omega t}
    \Tr
    \left[
      \bfG^a(n,\omega+n'\Omega) \bfSigma^b_\ell(n',\omega)
    \right]
    \\
    \label{eq:FTapp:2}
    \int dt' \int dt'' \Tr\left[\bfXi^{ab}_\ell(t,t'',t') \bfG^c(t',t'')\right]
    & =
    \sum_{nn'} \int \frac{d\omega}{2\pi} e^{-i(n+n')\Omega t}
    \Tr
    \left[
      \bfXi^{ab}_\ell(n,n',\omega) \bfG^c(n',\omega)
    \right]
  \end{align}
\end{subequations}
with
\begin{align}
  \label{eq:Xiab:FT}
  \bfXi^{ab}_\ell(n,n',\omega)
  \equiv
  \rcp\tau \int_0^\tau dt
  \int dt' \int dt'' \bfXi^{ab}_\ell(t,t'',t')
  e^{-i\omega(t'-t'')}
  e^{-in'\Omega (t'-t)}
  e^{in\Omega t}.
\end{align}
Importantly, the charge and heat current [see \eqnsref{eq:IcHT} and
(\ref{eq:HC:2}) together with \eqnref{eq:Ih:0}] are now expressed in the
frequency domain employing \eqnsref{eq:FTapp:1} and (\ref{eq:FTapp:2})
\textit{solely} in terms of the QD NEGF. This constitutes an \emph{exact} form
for the charge and heat currents of an interacting conductor coupled to
fermionic reservoirs.

One may want to interpret the above ac-driven current in terms of the
photo-assisted tunneling as done in the Tien-Gordon approach
\cite{Tien1963Jan}.  In this approach, the effect of the driving is transformed
to the appearance of the quasi-energy states ($\omega + n\Omega$) due to the
absorption/emission processes of photons and the resulting current is then
given by the sum of all possible processes, each of which is weighted by proper
Bessel functions with argument proportional to $\Psi_\ell$.
This simple interpretation does not work in our case though. The main reason is
that, while in the Tien-Gordon approach the time-dependent field is coupled to
the charge numbers ($c_{\ell\bfk\sigma}^\dag c_{\ell\bfk\sigma}$) only, in the
temperature-driven case it is coupled to the excitation energies
($\epsilon_{\ell\bfk}$) as well. Then, the photon-assisted processes with
different $n$ are no longer independent of each other but are instead
intermingled with each other and have very complicated energy dependencies [see
\eqnref{eq:SigmaLesser:exact}].  Hence, it is not practical to interpret the
temperature-driven current in terms of the photon-assisted processes.

Hereafter we adopt an approximation scheme using a
\textit{linear expansion in the Luttinger field} $\Psi_\ell(t)$. The reasons
are three-fold: (1) The Luttinger scheme was originally proposed to work only
in the linear response regime, (2) the second requirement of our Luttinger
setup, making the effective dot-contact coupling independent of $\Psi_\ell$, is
satisfied only in the linear response regime, and (3) in a practical manner the
use of the above exact form is limited because it is quite difficult to obtain
manageable analytical expressions for $\bfSigma^a_\ell(n,\omega)$ and
$\bfXi^{ab}_\ell(n,n',\omega)$ working for arbitrary magnitude of $\Psi_\ell$.

\subsection{Linear response regime}
\label{sec:linear_response_regime}

After deriving the general forms for the charge and heat currents in the
system, our intention is to supply a manageable formulation of such currents in
terms of the equilibrium or dynamical QD Green's functions. This objective is
reachable within the linear response regime, in which we treat the amplitudes
$\Psi_\ell$ as the smallest parameters and keep up to their \textit{linear
  order} in the charge and heat fluxes.

Considering that our periodic driving (\ref{eq:Psi}) is of the form,
$\Psi_\ell(t) = (\Psi_\ell/2) (e^{i\Omega t} + e^{-i\Omega t})$, in the linear
expansion with respect to $\Psi_\ell$, only the Fourier components with
$n,n'=0,\pm1$ of $\bfSigma^a_\ell(n,\omega)$ and $\bfXi^{ab}_\ell(n,n',\omega)$
should remain finite. While their explicit expressions and derivations can be
found in \appref{app:self_energies}, one should note that from
\eqnref{eq:SigmaRA} the linear expansion of $\Sigma^{R/A}_{\ell\sigma}(t,t')$
is given by
\begin{align}
  \bfSigma^{R/A}_\ell(t,t')
  \approx
  \mp i\bfGamma_\ell
  \left(
    1 + (2\lambda - 1) \Psi_\ell(t)
  \right)
  \delta(t - t')
\end{align}
so that $\bfSigma^{R/A}_\ell(t,t')$ becomes independent of the gravitational
field only when $\lambda = 1/2$. That is, only at this choice of $\lambda$, the
requirement that the dot-contact hybridization should be immune to the
temperature change is met.

In the same spirit of the linear expansion, only the $n=0,\pm1$ Fourier
components of the QD Green's functions should survive: $\bfG^a(0,\omega)$ is in
the zeroth order of $\Psi_\ell$, while $\bfG^a(\pm1,\omega)$ are linear in
$\Psi_\ell$: up to the linear order in $\Psi_\ell$,
\begin{align}
  \bfG^a(t,\omega)
  \approx
  \bfG^a(0,\omega)
  + e^{-i\Omega t} \bfG^a(1,\omega)
  + e^{+i\Omega t} \bfG^a(-1,\omega).
\end{align}
In particular, the $n=0$ components of the self energies and the QD Green's
functions should exactly correspond to their equilibrium values at
$\Psi_\ell=0$.

In the linear response regime, the charge and heat currents are described by
their first Fourier components only,
\begin{align}
  I^{c/h}_{\ell/\rm D}(t)
  =
  I^{c/h}_{\ell/\rm D}(\Omega) e^{-i\Omega t}
  +
  I^{c/h}_{\ell/\rm D}(-\Omega) e^{+i\Omega t}
\end{align}
with $I^{c/h}_{\ell/\rm D}(-\Omega) = [I^{c/h}_{\ell/\rm D}(\Omega)]^*$ since
no current flows at $\Psi_\ell=0$.  In the following sections, we are going to
express the Fourier components $I^{c/h}_\ell(\Omega)$ of the charge and heat
currents in terms of the equilibrium QD Green's functions, $\bfG^a(0,\omega)$
and the nonequilibrium linear components, $\bfG^a(\pm1,\omega)$.

\subsection{Charge/Heat currents and power}

In order to obtain the charge current in the linear response regime, we express
the contact charge current (\ref{eq:Ic:2}) in terms of the Fourier components
$\bfG^a(n,\omega)$ and $\bfSigma^a_\ell(n,\omega)$ by using \eqnref{eq:FTapp:1}
and then keep only the linear-order terms in $\Psi_\ell$ [refer to
\appref{app:self_energies}]. The derivation is quite straightforward and we
obtain
\begin{align}
  \label{eq:Ic:linear}
  \begin{split}
    I^c_\ell(\Omega)
    & =
    e
    \int \frac{d\omega}{2\pi}
    \Tr
    \bigg[
    (2i\bfGamma_\ell)
    \bigg(
    - \frac{\Psi_\ell}{2}
    \Delta_f(\omega+\Omega,\omega)
    \left(\omega + \frac{\Omega}{2}\right)
    \left(
      \bfG^R(0,\omega+\Omega)
      -
      \bfG^A(0,\omega)
    \right)
    \\
    & \qquad\qquad\qquad\qquad\qquad\mbox{}
    +
    f(\omega+\Omega)
    \left(
      \bfG^R(1,\omega+\Omega)
      -
      \bfG^A(1,\omega)
    \right)
    +
    \bfG^<(1,\omega)
    \bigg)
    \bigg]
  \end{split}
\end{align}
and
\begin{align}
  \label{eq:ID:linear}
  I^c_{\rm D}(\Omega)
  =
  e \int \frac{d\omega}{2\pi}
  \Omega \Tr \bfG^<(1,\omega),
\end{align}
where we have defined
\begin{align}
  \label{eq:DeltaF}
  \Delta_f(\omega,\omega')
  \equiv
  \frac{f(\omega) - f(\omega')}{\omega-\omega'}
\end{align}
[refer to \appref{sec:linear_response_regime:currents} for details]. As expected, the charge current depends not only on the
equilibrium QD Green's functions but also on the dynamical ones,
$\bfG^{R/A/<}(1,\omega)$, even though the linear response ($\Psi_\ell\to0$) is
taken. It is obviously because our perturbations are dynamical and the
dynamical excitations of the system, even though being small, cannot be
described solely in terms of the equilibrium Green's functions.

Now we turn to the heat transport. In order to find the expressions for the
heat currents, the power, and the energy change rates in the linear response
regime, we write down the energies (\ref{eq:HT:2}) and (\ref{eq:HC:2}) in terms
of the Fourier components of $\bfG^a(n,\omega)$, $\bfSigma^a_\ell(n,\omega)$,
and $\bfXi^{ab}_\ell(n,n',\omega)$ by using \eqnsref{eq:FTapp:1} and
(\ref{eq:FTapp:2}) and then keep only the linear-order terms in $\Psi_\ell$
[see \appref{app:self_energies} for details]. Following the explicit derivation
in \appref{sec:linear_response_regime:currents}, the contact heat currents can
be explicitly obtained as
\begin{align}
  \label{eq:Ih:linear:final}
  \begin{split}
    I^h_\ell(\Omega)
    & =
    \hbar
    \int \frac{d\omega}{2\pi}
    \Tr
    \bigg[
    (2i\bfGamma_\ell)
    \bigg(
    \frac{\Psi_\ell}{2}
    \Delta_f(\omega+\Omega,\omega)
    \left(\omega + \frac{\Omega}{2}\right)^2
    \left(
      \bfG^R(0,\omega+\Omega) - \bfG^A(0,\omega)
    \right)
    \\
    & \qquad\quad\mbox{}
    -
    \left(\omega+\frac{\Omega}{2}\right)
    \left(
      f(\omega)
      \bfG^R(1,\omega)
      -
      f(\omega+\Omega)
      \bfG^A(1,\omega)
      +
      \bfG^<(1,\omega)
    \right)
    \bigg)
    \bigg] + I^h_{\ell\rm T}(\Omega),
  \end{split}
\end{align}
where
$I^h_{\ell\rm T}(\Omega) \equiv - \frac{\Psi_\ell}{2} \frac{i\Omega}{4} E_{\rm
  T\ell0}$ is an additional unphysical term.  \Eqnsref{eq:Ic:linear} and
(\ref{eq:Ih:linear:final}) are our main results.

The expression given in (\ref{eq:Ih:linear:final}) for the contact heat current
needs a discussion. The additional term $I^h_{\ell\rm T}(\Omega)$ which is
proportional to $E_{\rm T\ell0}$ is an \textit{artefact} of our Luttinger's
trick. In our setup, we dynamically drive the contact by the field
$\Psi_\ell(t)$ and the tunneling barrier by the field $\lambda\Psi_\ell(t)$ so
that an effective energy capacitor which is dynamically driven by the field
difference $(1-\lambda) \Psi_\ell(t)$ is formed. However, this effect is not
contained in the original system and is solely due to the Luttinger's setup
itself. This artificial setup then gives rise to an additional heat transfer
$(1-\lambda) \Psi_\ell(t) E_{\rm T\ell0}$ (up to the linear order) between the
contact $\ell$ and the tunneling barrier, resulting in
$I^h_{\ell\rm T}(\Omega)$ in the contact heat current. In fact, in the next
section for the noninteracting system, we compare the results from our
Luttinger's trick and those obtained from the scattering theory based on the
dissipation-fluctuation theorem and find that two results are identical only
when this artefact is not taken into account. Therefore, we will drop out the
term $I^h_{\ell\rm T}(\Omega)$ from now on.

Finally, we find the expression for the power of dissipation (\ref{eq:P:0}) in
the linear response regime [see \eqnref{eq:P:linear}]. In particular, we focus
on the time average of the power which is simplified to
\begin{align}
  \label{eq:P}
  \overline{P}
  = -\sum_\ell \Psi_\ell \re[I^h_\ell(\Omega)].
\end{align}
As expected, it reflects that the time-averaged power is directly related to
the real part of the Fourier component of the contact heat currents. It is
because the dissipation happens at the contacts and the time average picks up
only the dissipative effect.

\subsection{Application of conservation law and sum rule}

Note that our expressions for the charge and heat currents,
\eqnsref{eq:Ic:linear} and (\ref{eq:Ih:linear:final}) necessitate the knowledge
of the nonequilibrium components $\bfG^<(1,\omega)$ as wells as
$\bfG^{R/A}(1,\omega)$.  Technically, it is much harder to theoretically obtain
the lesser Green's functions than the retarded/advanced Green's functions
because the lesser ones reflects not only the energy excitations of the system
but also their nonequilibrium distribution. Therefore, our expressions for the
currents would be more useful if the knowledge of the lesser Green's functions
is avoided, especially when the quantum dot is interacting.

We have found that it is possible as long as the hybridization matrix
$\bfGamma_\ell$ is proportional to the identity matrix:
\begin{align}
  \label{eq:simplifying_condition}
  \bfGamma_\ell = \Gamma_\ell {\bf 1}
\end{align}
in which condition,
$\Tr[\bfGamma_\ell \bfG^<(1,\omega)] = \Gamma_\ell \Tr[\bfG^<(1,\omega)]$. This
condition cannot accommodate the general situations in real experiments, but
most of qualitative features can be still captured within this condition. So
this simplifying condition is acceptable at least for theoretical studies.

First, the expression for the contact charge current, \eqnref{eq:Ic:linear} can
be simplified further if the charge conservation is taken into account. If we
apply the charge conservation (\ref{eq:chargeconservation}) up to the linear
order, then we have
\begin{align}
  \label{eq:charge_conservation:linear}
  \sum_\ell I^c_\ell(\Omega) + I^c_d(\Omega) = 0
\end{align}
which enables one to write down the integral
$\int d\omega\, \Tr[\bfG^<(1,\omega)]$ in terms of the other components of the
QD Green's functions [see \eqnref{eq:QDGLesser:linear:cc:final}]. Then, we can
get a nice expression for the \emph{interacting} charge current at the contact
$\ell$ in the linear response regime which writes \emph{solely} in terms of the
retarded/advanced QD Green's functions:
\begin{align}
  \label{eq:Ic:linear:final}
  \begin{split}
    I^c_\ell(\Omega)
    & =
    e
    \int \frac{d\omega}{2\pi}
    \Bigg[
    \left(
      \sum_{\ell'}
      \frac{\Psi_{\ell'}}{2} \frac{2i\Gamma_{\ell'}}{2i\Gamma + \Omega}
      -
      \frac{\Psi_\ell}{2}
    \right)
    \\
    & \qquad\qquad\qquad\qquad\quad\mbox{}
    \times
    \Delta_f(\omega+\Omega,\omega)
    \left(\omega + \frac{\Omega}{2}\right)
    (2i\Gamma_\ell)
    \Tr
    \left[
      \bfG^R(0,\omega+\Omega)
      -
      \bfG^A(0,\omega)
    \right]
    \\
    & \qquad\qquad\qquad\mbox{}
    +
    \frac{\Omega}{2i\Gamma+\Omega} f(\omega+\Omega) (2i\Gamma_\ell)
    \Tr
    \left[
      \bfG^R(1,\omega+\Omega)
      -
      \bfG^A(1,\omega)
    \right]
    \Bigg]
  \end{split}
\end{align}
with
\begin{align}
  \Gamma \equiv \sum_\ell \Gamma_\ell.
\end{align}

The expression of the contact heat current, \eqnref{eq:Ih:linear:final} can be
also further simplified in a similar way. Recall that we have the sum rule
(\ref{eq:Wsum}) for the energy change rates, which gives rise to
\begin{align}
  \label{eq:Wsum:2}
  \sum_\ell (W_{\rm C\ell}(\Omega) + W_{\rm T\ell}(\Omega)) + W_{\rm D}(\Omega) = 0
\end{align}
in the linear response regime. This sum rule enables us to replace the integral
$\int d\omega\, \omega \Tr[\bfG^<(1,\omega)]$ with ones of other QD Green's
functions. However, this sum rule cannot be constructed without knowing
explicitly the form of the QD Hamiltonian $\varH_{\rm D}$ since the change rate
$W_{\rm D}(\Omega)$ depends on $\varH_{\rm D}$.
Therefore, our strategy is as follows: Once $\varH_{\rm D}$ is known, we find
out the expression for $W_{\rm D}(\Omega)$ in terms of the QD NEGFs in the
linear response regime and then write down the integral
$\int d\omega\, \omega \Tr[\bfG^<(1,\omega)]$ in terms of other QD NEGFs by
using the sum rule (\ref{eq:Wsum:2}) and the explicit expression for the
partial sum, $\sum_\ell (W_{\rm C\ell}(\Omega) + W_{\rm T\ell}(\Omega))$ [see
\eqnref{eq:Wsumrule:WW} in \appref{sec:linear_response_regime:currents}].


In the following sections, we apply our formalism to two specific examples: the
noninteracting case and the interacting case with the Hartree
approximation. Especially, in the noninteracting case, we demonstrate the
justification of the choice $\lambda = 1/2$ in more details.

\section{Noninteracting Case}
\label{sec:ni}

As a first application of our formalism derived in the previous sections, we
consider the noninteracting case in which the QD Hamiltonian (\ref{eq:HD}) is
given by
\begin{align}
  \label{eq:HD:ni}
  \varH_{\rm D}
  = \sum_\sigma \epsilon_\sigma d_\sigma^\dag d_\sigma.
\end{align}
We assume that the dot-lead coupling is spin-independent so that $\bfGamma$,
now being $2\times2$ matrix, is diagonal and given by
$\bfGamma = \Gamma {\bf 1}$. The orbital index $m$ is now replaced by the spin
index $\sigma$, and the trace over $m$ is now the summation over $\sigma$.
It is then quite straightforward to derive and solve the Dyson's equations for
the QD NEGFs, $\varG^{R/A}_\sigma(t,t')$ and to find out their Fourier
components in the linear response regime [refer to \appref{app:ni} for detailed
derivations]. The $n=0$ (equilibrium) component of the retarded/advanced QD
Green's functions are found to be
\begin{align}
  \label{eq:QDGF:ni}
  \varG^{R/A}_\sigma(0,\omega)
  = \rcp{\omega - \epsilon_\sigma/\hbar \pm i \Gamma}
\end{align}
and their $n=1$ components exactly vanish at $\lambda = 1/2$, that is,
$\varG^{R/A}_\sigma(\pm1,\omega) = 0$ since
$\Sigma^{R/A}_{\ell\sigma}(\pm1,\omega) = 0$.  On the other hand,
$\varG^<_\sigma(\pm1,\omega)$ does not vanish at $\lambda = 1/2$, so the
dynamical components of the QD Green's functions are still relevant in
time-dependent charge and heat transports. However, as explained in the
previous section, we do not seek out the explicit expression for
$\varG^<_\sigma(\pm1,\omega)$, but instead resort to the charge conservation
(\ref{eq:charge_conservation:linear}) and the sum rule (\ref{eq:Wsum:2}) to
derive the integrals of $\varG^<_\sigma(\pm1,\omega)$.

Starting from the general formulas of the charge and heat currents,
\eqnsref{eq:Ic:linear:final} and (\ref{eq:Ih:linear:final}), one can derive the
explicit expressions for the charge and heat currents [see \eqnsref{eq:Ic:ni}
and (\ref{eq:Ih:ni})] for the noninteracting case, in terms of the
\textit{thermoelectric admittances} $L_{\ell\ell'}(\Omega)$ and the
\textit{thermal admittances} $K_{\ell\ell'}(\Omega)$:
\begin{align}
  I^c_\ell(\Omega)
  = \sum_{\ell'} L_{\ell\ell'}(\Omega) \frac{\Psi_{\ell'}}{2},
  \
  I^h_\ell(\Omega)
  = \sum_{\ell'} K_{\ell\ell'}(\Omega) \frac{\Psi_{\ell'}}{2},
\end{align}
where the diagonal components are decomposed into
$L_{\ell\ell}(\Omega) \equiv L_\ell(\Omega) - L_{\ell\bar\ell}(\Omega)$ and \linebreak
$K_{\ell\ell}(\Omega) \equiv K_\ell(\Omega) - K_{\ell\bar\ell}(\Omega)$. The
self admittances are then found to be
\begin{subequations}
  \label{eq:LK:ni:self}
  \begin{align}
    L_\ell(\Omega)
    & = (2i\Gamma_\ell)e \Omega  \sum_\sigma P_{1\sigma}(\Omega),
    \\
    K_\ell(\Omega)
    & = (2i\Gamma_\ell)(-\hbar) \Omega  \sum_\sigma P_{2\sigma}(\Omega),
  \end{align}
\end{subequations}
and the cross admittances are given by
\begin{subequations}
  \label{eq:LK:ni:cross}
  \begin{align}
    L_{\ell\bar\ell}(\Omega)
    & = L_{\bar\ell\ell}(\Omega)
    = 4 \Gamma_{\Left} \Gamma_{\Right} e \sum_\sigma P_{1\sigma}(\Omega),
    \\
    K_{\ell\bar\ell}(\Omega)
    & = K_{\bar\ell\ell}(\Omega)
    = 4 \Gamma_{\Left} \Gamma_{\Right} (-\hbar) \sum_\sigma P_{2\sigma}(\Omega).
  \end{align}
\end{subequations}
where we have defined
\begin{align}
  \label{eq:Pn}
  P_{n\sigma}(\omega)
  \equiv
  \int \frac{d\omega}{2\pi}
  \Delta_f(\omega+\Omega,\omega)
  \left(\omega+\frac{\Omega}{2}\right)^n
  \varG^R_\sigma(0,\omega+\Omega)
  \varG^A_\sigma(0,\omega).
\end{align}



\begin{figure}[t]
  \centering \includegraphics[width=0.5\textwidth]{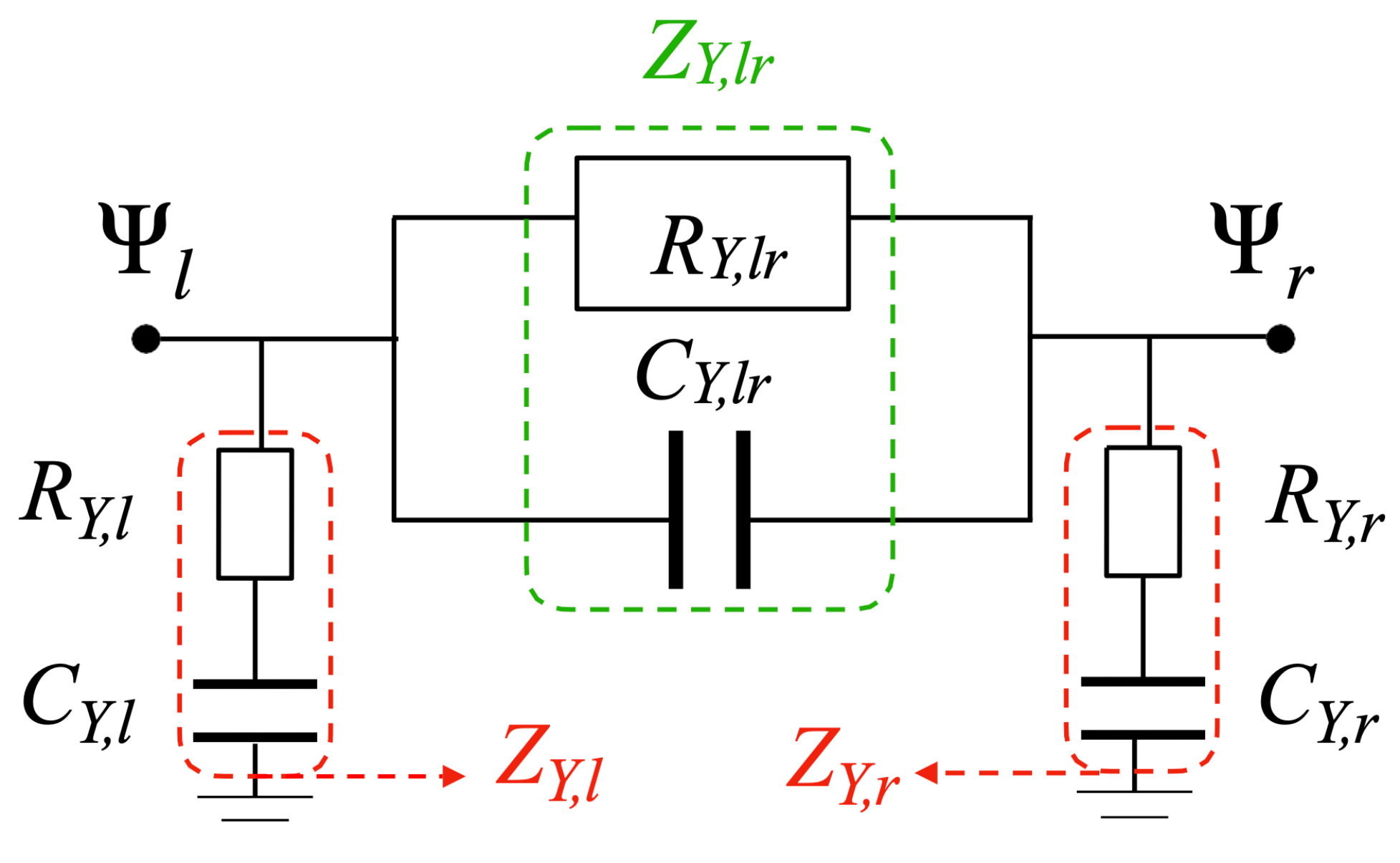}
  \caption{Equivalent RC circuit in the low temperature and low frequency
    limits. The RC circuit describes both the thermoelectrical (with $Y = L$)
    and the thermal admittance (with $Y = K$) in terms of resistors
    $R_{Y,\ell}$, capacitors $C_{Y,\ell}$ and the cross resistance and
    capacitances $R_{Y,\Left\Right}$, and $C_{Y,\Left\Right}$. Refer to the
    main text for their specifications for either the thermoelectrical or the
    thermal transports.}
  \label{fig:RCcircuit}
\end{figure}

These linear-response admittances can be expressed in terms of an equivalent RC
circuit, as shown in \figref{fig:RCcircuit}. Under this equivalence, the cross
admittances $L_{\Left\Right}$ and $K_{\Left\Right}$ represent the thermoelectric ($Y=L$)
and thermal conductance ($Y=K$) between the two contacts, coming from the
\emph{parallel} configuration of a resistor $R_{Y,\Left\Right}$ and a capacitor
$C_{Y,\Left\Right}$ so that
\begin{align}
  \label{eq:RC:cross}
  Y_{\Left\Right}(\Omega)
  = \rcp{Z_{Y,\Left\Right}(\Omega)}
  = \rcp{R_{Y,\Left\Right}} + \rcp{1/i\Omega C_{Y,\Left\Right}}.
\end{align}
On the other hand, the self admittances $L_\ell$ and $K_\ell$ represent the
electrical/thermal conductances for charging/heating and relaxing in the
contact $\ell$, coming from the \emph{serial} configurations of a resistor
$R_{Y,\ell}$ and a capacitor $C_{Y,\ell}$ so that
\begin{align}
  Y_\ell(\Omega)
  = \rcp{Z_{Y,\ell}(\Omega)}
  = \rcp{\displaystyle R_{Y,\ell} + 1/i\Omega C_{Y,\ell}}.
\end{align}
Generally, the above resistances and capacitances are functions of the
frequency $\Omega$.  However, in the low-frequency ($\hbar\Omega \ll k_BT$)
limit, they can be approximated to constants. We also assume the
low-temperature ($k_BT \ll \Gamma_\ell$) condition to apply the Sommerfeld
approximation. The self resistances and capacitances are then approximated to
\begin{subequations}
  \label{eq:L:self:ni:lowfreq}
  \begin{align}
    C_{L,\ell}
    & = -eh G_{\rm th} T \frac{\Gamma_\ell}{\Gamma} \sum_\sigma \rho'_\sigma(0),
    \\
    R_{L,\ell}
    & =
    \rcp{e G_{\rm th} T} \frac{\Gamma}{\Gamma_\ell}
    \frac{\sum_\sigma \rho'_\sigma(0) \rho_\sigma(0)}%
    {[\sum_\sigma \rho'_\sigma(0)]^2},
  \end{align}
\end{subequations}
and
\begin{subequations}
  \label{eq:K:self:ni:lowfreq}
  \begin{align}
    C_{K,\ell}
    & = h G_{\rm th} T \frac{\Gamma_\ell}{\Gamma} \sum_\sigma \rho_\sigma(0),
    \\
    R_{K,\ell}
    & =
    -
    \rcp{2G_{\rm th} T} \frac{\Gamma}{\Gamma_\ell}
    \frac{\sum_\sigma [\rho_\sigma(0)]^2}%
    {[\sum_\sigma \rho_\sigma(0)]^2},
  \end{align}
\end{subequations}
where $G_{\rm th} \equiv \frac{\pi^2}{3} \frac{k_B^2T}{h}$ is the thermal
relaxation resistance for a single mode mesoscopic capacitor.
On the other hand, the low-frequency cross resistances and capacitances are
found to
\begin{subequations}
  \label{eq:L:cross:ni:lowfreq}
  \begin{align}
    R_{L,\Left\Right}
    & =
    \rcp{(-e)hG_{\rm th} T}
    \frac{\Gamma}{2\Gamma_{\Left}\Gamma_{\Right}}
    \rcp{\sum_\sigma \rho'_\sigma(0)},
    \\
    C_{L,\Left\Right}
    & =
    (-e)h^2G_{\rm th} T
    \frac{2\Gamma_{\Left}\Gamma_{\Right}}{\Gamma}
    \sum_\sigma \rho'_\sigma(0) \rho_\sigma(0),
  \end{align}
\end{subequations}
and
\begin{subequations}
  \label{eq:K:cross:ni:lowfreq}
  \begin{align}
    R_{K,\Left\Right}
    & =
    \rcp{hG_{\rm th} T}
    \frac{\Gamma}{2\Gamma_{\Left}\Gamma_{\Right}}
    \rcp{\sum_\sigma \rho_\sigma(0)},
    \\
    C_{K,\Left\Right}
    & =
    \frac{h^2G_{\rm th} T}{2}
    \frac{2\Gamma_{\Left}\Gamma_{\Right}}{\Gamma}
    \sum_\sigma [\rho_\sigma(0)]^2.
  \end{align}
\end{subequations}
The detailed analysis of the resistances and capacitances will be present in
\secref{sec:ni:analysis}.

\subsection{Why the choice of $\lambda = 1/2$ ?}

In the original works \cite{Ludovico2014apr,Ludovico2016} which explored the
heat transport in time domain, it is found that the meaningful definition of
the contact energy should be determined to be $\varQ_\ell(t)$ with
$\lambda = 1/2$ [see \eqnref{eq:Q:0}] because only this choice is consistent
with the first and second laws of thermodynamics.
While this argument alone justifies the
choice of $\lambda = 1/2$, in this section we intend to find more evidence
which supports the choice of $\lambda = 1/2$ by comparing our results for
$I^c_\ell(\Omega)$ and $I^h_\ell(\Omega)$ with the previous ones obtained by
different methods for the similar systems.


By using the equation-of-motion method, Rossell\'{o}, L\'{o}pez, and Lim
\cite{Rossello2015} have investigated the dynamical heat current through the
similar setup as ours but with a single contact which is driven by an ac
electric voltage. In calculating the heat current, they also took into account
the energy barrier contribution with $\lambda = 1/2$, as proposed in
Ref.~\cite{Ludovico2014apr}.  As expected, our self thermoelectric admittance
$L_\ell(\Omega)$ is exactly equal to their self electrothermal admittance
$M_\ell(\Omega)/T$ [see Eq.~(36) in Ref.~\cite{Rossello2015}] which measures
the heat current through the contact with respect to the ac electric driving in
the contact. It justifies that the dynamical gravitational field in our setup
should be coupled to $\varQ_\ell$ with $\lambda = 1/2$ in order to get the
correct dynamical charge current.  Note that this agreement,
$M_\ell(\Omega) = T L_\ell(\Omega)$ (where $T$ is the background temperature,
the common temperature in the contacts) also reflects the fact that the
\emph{reciprocal relation}, or the so-called Onsager's relation
\cite{Onsager1931} should hold in the thermal transport. However, in the point
of view of the fluctuation-dissipation theorem, both admittances
$L_\ell(\Omega)$ and $M_\ell(\Omega)$ in the linear response regime are related
to the same fluctuation
\begin{equation}
  \Avg{\comm{\ode{\varN_\ell}{t}}{\ode{\varQ_\ell}{t}}}.
\end{equation}
In our setup, the perturbative (gravitational) field is coupled to $\varQ_\ell$
and $\avg{d\varN_\ell/dt}$ is measured, while in Rossell\'{o}'s work the
external (electric) field is coupled to $\varN_\ell$ and $\avg{d\varQ_\ell/dt}$
is measured. So, apparently, the agreement might be mathematically trivial
because both used the same $\varQ_\ell$ with $\lambda = 1/2$.

The second previous work to be compared with is the work done by Lim,
L\'{o}pez, and S\'{a}nchez \cite{Lim2013nov} which has applied the
\emph{scattering theory} to the single-contact QD setup to obtain the dynamical
charge and heat current in the linear response regime when the contact is
driven either by an ac voltage or by an ac temperature. Note that in the
scattering theory approach the barrier plays the role of the scatterer only, so
no energy is stored in it and the contact energy is defined with respect to
$\varH_\ell$ only. They calculated the low-frequency responses of the currents
with respect to the ac voltage and temperature, via the fluctuation-dissipation
relation which they assumes to hold. We have found that our low-frequency
expansion of the self admittances, \eqnsref{eq:L:self:ni:lowfreq} and
(\ref{eq:K:self:ni:lowfreq}) are in good agreement with the scattering-theory
predictions [see Eqs.~(7), (8), and (9) in
Ref.~\cite{Lim2013nov}]. This agreement strongly justifies our use of
$\lambda = 1/2$, especially because they are derived from two different
approaches: We have directly calculated the dissipative part by adopting the
Luttinger's trick, while in Ref.~\cite{Lim2013nov} the admittances were
obtained by calculating the fluctuations based on the scattering theory. Also,
this agreement implies that the fluctuation-dissipation theorem holds for
thermal transport, at least in the non-interacting and single-contact case.


\subsection{More analysis on charge/heat currents}
\label{sec:ni:analysis}

\begin{figure*}[t]
  \centering
  \includegraphics[width=7cm]{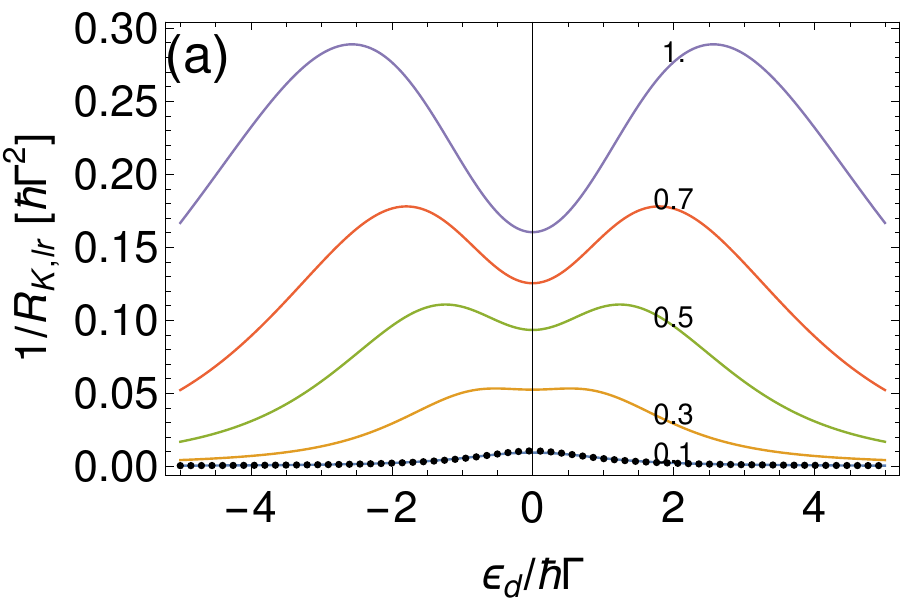}%
  \includegraphics[width=7cm]{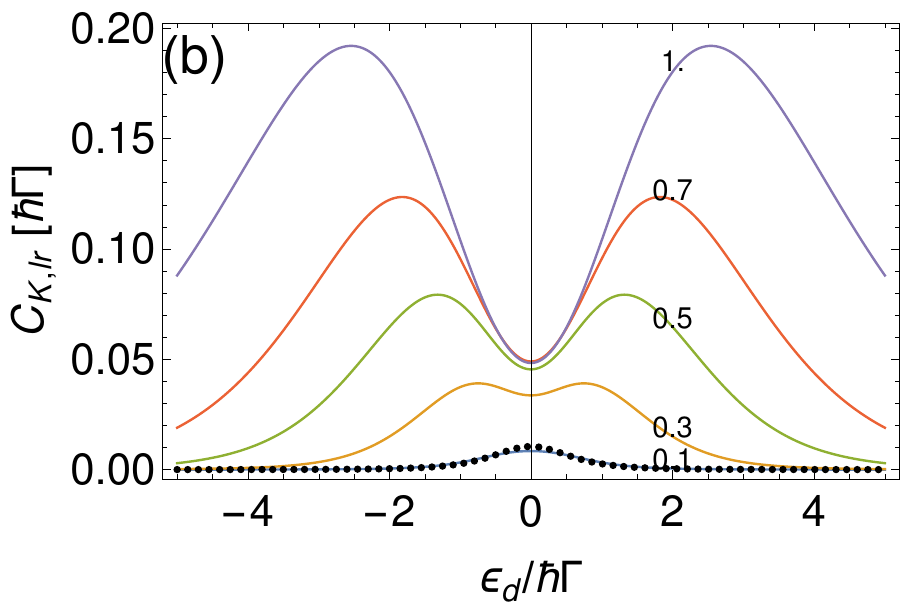}\\
  \includegraphics[width=7cm]{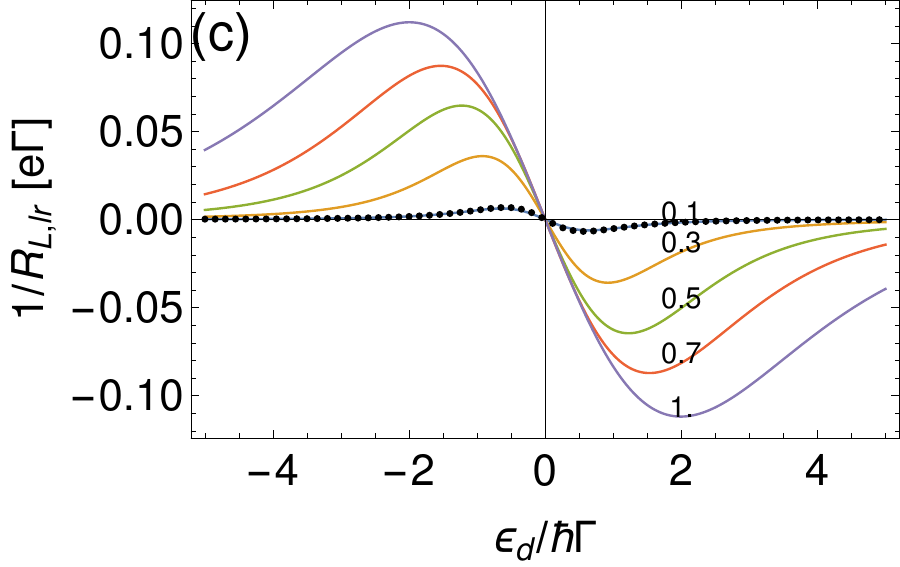}%
  \includegraphics[width=7cm]{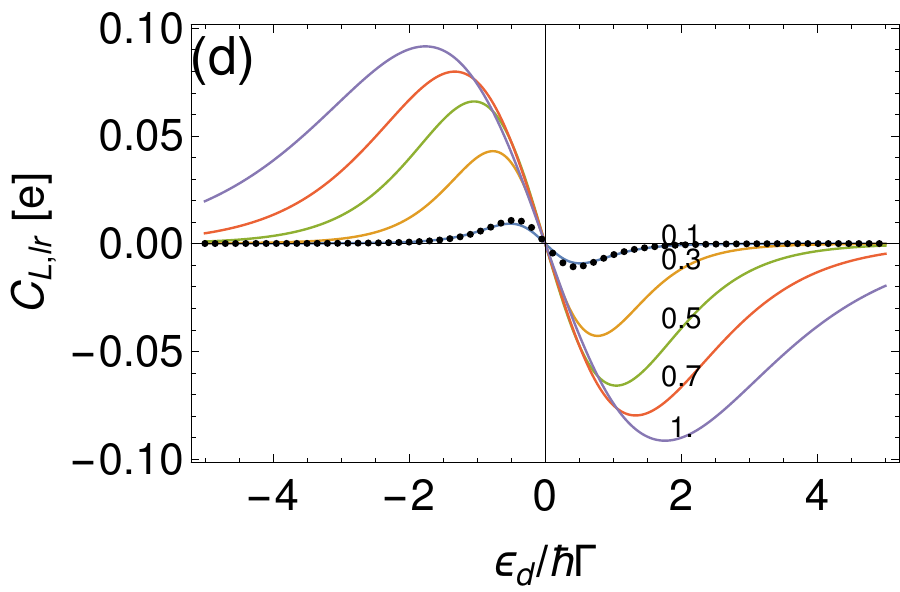}
  \caption{(a) The cross thermal conductance $1/R_{K,\Left\Right}$ and (b) the
    cross thermal capacitance $C_{K,\Left\Right}$, and (c) the cross
    thermoelectric conductance $1/R_{L,\Left\Right}$ and (d) the cross
    thermoelectric capacitance $C_{L,\Left\Right}$ in the low-frequency limit
    as functions of the spin-degenerate dot level $\epsilon_d$ in the
    noninteracting case with the symmetric coupling,
    $\Gamma_{\Left} = \Gamma_{\Right}$.  Each curve corresponds to the
    different temperature $k_BT/\hbar\Gamma$ whose value is annotated.  The
    resistances and capacitors for the cross admittances are defined via
    \eqnref{eq:RC:cross} and the corresponding cross thermal and thermoelectric
    admittances are evaluated via \eqnref{eq:LK:ni:cross} by using given values
    of temperatures and the value of frequency chosen numerically as small as
    possible.  The black dotted lines correspond to the low-temperature limits
    as given by \eqnsref{eq:L:cross:ni:lowfreq} and
    (\ref{eq:K:cross:ni:lowfreq}).}
  \label{fig:ni:LK:cross}
\end{figure*}

\begin{figure*}[t]
  \centering
  \includegraphics[width=7cm]{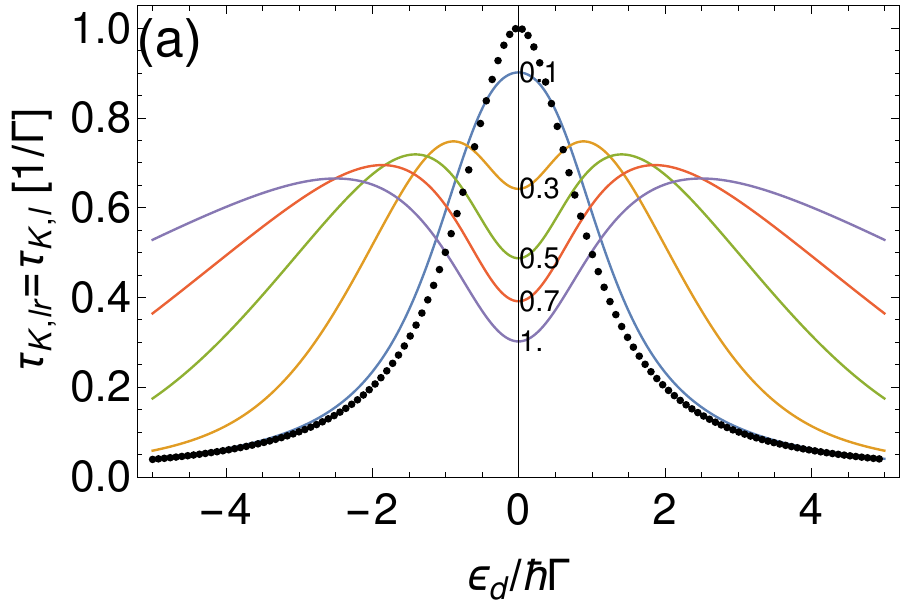}%
  \includegraphics[width=7cm]{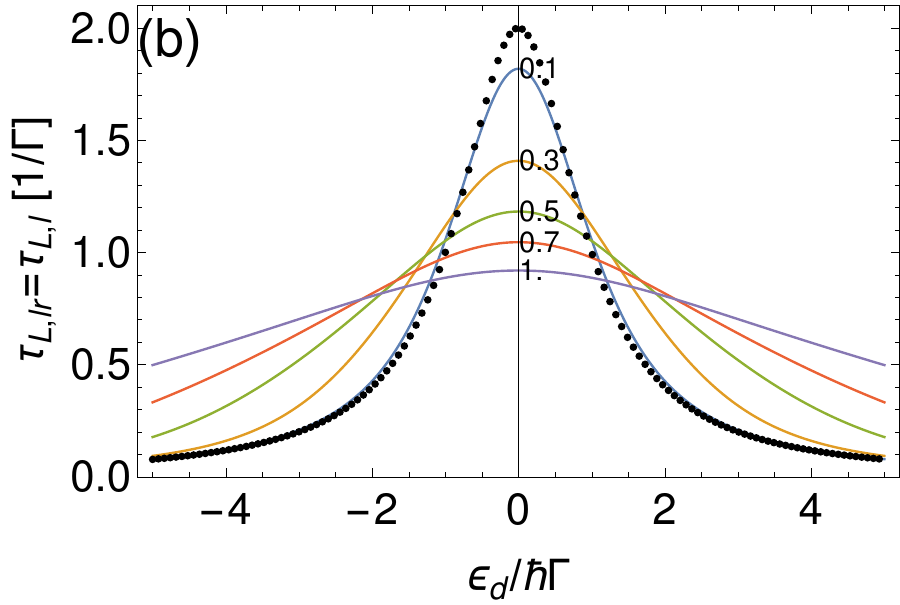}
  \caption{(a) The RC times $\tau_{K,\Left\Right} = \tau_{K,\ell}$ from thermal
    admittance and (b) the RC time $\tau_{L,\Left\Right} = \tau_{L,\ell}$ from
    thermoelectric admittance in the low-frequency limit as functions of the
    spin-degenerate dot level $\epsilon_d$ in the noninteracting case with the
    same parameters as in \figref{fig:ni:LK:cross}.  The black dotted lines
    correspond to the low-temperature limits as given by
    \eqnref{eq:RCtime:ni:lowfreq:lowtemp}.}
  \label{fig:ni:LK:time}
\end{figure*}

Since the physical discussion on the low-frequency self admittances,
$R_{L/K,\ell}$ and $C_{L/K,\ell}$ have already been done in
Ref.~\cite{Lim2013nov}, we focus on the cross admittances here.
First, the low-frequency and low-temperature expansions of the cross
resistances $R_{L/K,\Left\Right}$ are physically reasonable and are in agreement
with the scattering-theory prediction. For example, the low-temperature cross
thermal conductance $1/R_{K,\Left\Right}$ [see the black dotted line in
\figref{fig:ni:LK:cross}~(a)] was predicted to be proportional to the electric
conductance
$(e^2/h) (2\Gamma_{\Left}\Gamma_{\Right}/\Gamma) \sum_\sigma \rho_\sigma(0)$
\cite{Averin2010}, which is well reflected in \eqnref{eq:K:cross:ni:lowfreq},
while the low-temperature cross thermoelectric conductance $1/R_{L,\Left\Right}$
[see the black dotted line in \figref{fig:ni:LK:cross}~(c)] is proportional to
$\sum_\sigma \rho_\sigma'(0)$ as expected.
It should be noted that the fluctuation-dissipation theorem applied to the heat
transport through two-contact systems is no longer valid because scattering
events that connect two different terminals induce a nonvanishing term for the
equilibrium heat-heat correlation function at the low temperature limit, which
is incompatible with the expected behavior of $K_{\Left\Right}(\Omega)$
\cite{Averin2010,Sergi2011}. Therefore, one cannot exploit the
fluctuation-dissipation theorem to study the dynamic heat transport through the
quantum-dot systems described in terms of the tight-binding model. It signifies
that our Luttinger formalism is the promising candidate for the systematic
study of dynamical temperature driving.

\Figref{fig:ni:LK:cross} displays the dot-level dependencies of the
thermal/thermoelectric conductances and capacitances for a wide range of
temperatures. The thermoelectric conductance $1/R_{L,\Left\Right}$ and capacitances
$C_{L,\Left\Right}$ share similar dependence on the dot level, whose qualitative
feature does not change much as the temperature increases [see
\figsref{fig:ni:LK:cross}~(c) and (d)].
While the thermal conductance $1/R_{K,\Left\Right}$ and capacitances $C_{K,\Left\Right}$
also share similar dependence on the dot level, it changes from single-peak
shape to double-peak one as the temperature increases [see
\figsref{fig:ni:LK:cross}~(a) and (b)]. Recall that the heat current depends
not only on the carrier occupation but also the carrier energy. At high
temperatures, high-energy carrier can make more contribution to the heat
current, which is the reason why the heat current can be larger at the
off-resonant condition, as demonstrated in \figsref{fig:ni:LK:cross}~(a) and (b).

Very interesting property peculiar to the noninteracting condition can be found
in the RC times defined as
\begin{align}
  \tau_{Y,\ell} \equiv |R_{Y,\ell} C_{Y,\ell}|
  \quad\text{and}\quad
  \tau_{Y,\Left\Right} \equiv |R_{Y,\Left\Right} C_{Y,\Left\Right}|
\end{align}
for $Y = L,K$. In the noninteracting case, the self and cross RC times are
always equal to each other, that is,
\begin{align}
  \tau_{Y,\Left} = \tau_{Y,\Right} = \tau_{Y,\Left\Right}
\end{align}
for both $Y = L,K$. It is because the self and cross admittances share the same
frequency dependence: From \eqnsref{eq:LK:ni:self} and (\ref{eq:LK:ni:cross}),
one can find that $L_\ell(\Omega)$ and $L_{\Left\Right}(\Omega)$ are proportional to
$\sum_\sigma P_{1\sigma}(\Omega)$, while $K_\ell(\Omega)$ and
$K_{\Left\Right}(\Omega)$ are proportional to $\sum_\sigma P_{2\sigma}(\Omega)$. As
we will see in the next section, this is not the case for interacting QD
systems. That is, the comparison between the self and cross response times can
be used to measure the effect of the interaction. \Figref{fig:ni:LK:time} shows
the dot-level dependence of the RC times. The black dotted lines correspond to
the low-temperature limit which is given by
\begin{align}
  \label{eq:RCtime:ni:lowfreq:lowtemp}
  \tau_{L,\Left\Right} = 2\tau_{K,\Left\Right}
  = h \frac{\sum_\sigma [\rho_\sigma(0)]^2}{\sum_\sigma \rho_\sigma(0)}.
\end{align}
While $\tau_{L,\Left\Right}$ shows the behavior similar to the QD density of states
which broadens as the temperature increases, $\tau_{K,\Left\Right}$ features the
double-peak structure at high temperatures.

Finally, the time-averaged power (\ref{eq:P}) for the noninteracting case is
obtained as
\begin{align}
  \label{eq:P:ni}
  \overline{P}(\Omega)
  =
  - \sum_\ell \frac12 \Psi_\ell^2 \Omega^2 C_{K,\ell}^2 R_{K,\ell}
  + \frac12 \frac{(\Psi_{\Left} - \Psi_{\Right})^2}{R_{K,\Left\Right}}.
\end{align}
We then found that in the low-frequency limit only the second term remains
finite so that only the cross thermal resistance $R_{K,\Left\Right}$ is responsible
for the energy dissipation. Interestingly, this second term for the dissipation
is identical to its electric counterpart, $V^2/2R$ where $V$ is the electric
voltage drop and $R$ is the electric resistance between the two contacts.
We again would like to stress that \eqnref{eq:P:ni} is a natural outcome
obtained by following the procedure based on our Luttinger formalism, without
resorting to some heuristic arguments. Therefore our formalism is proven to
provide a systematic way to investigate the dynamical heat transport in the
linear response regime.

\section{Interacting Case: Hartree Approximation}
\label{sec:hartree}

\begin{figure*}[!t]
  \centering
  \includegraphics[width=7cm]{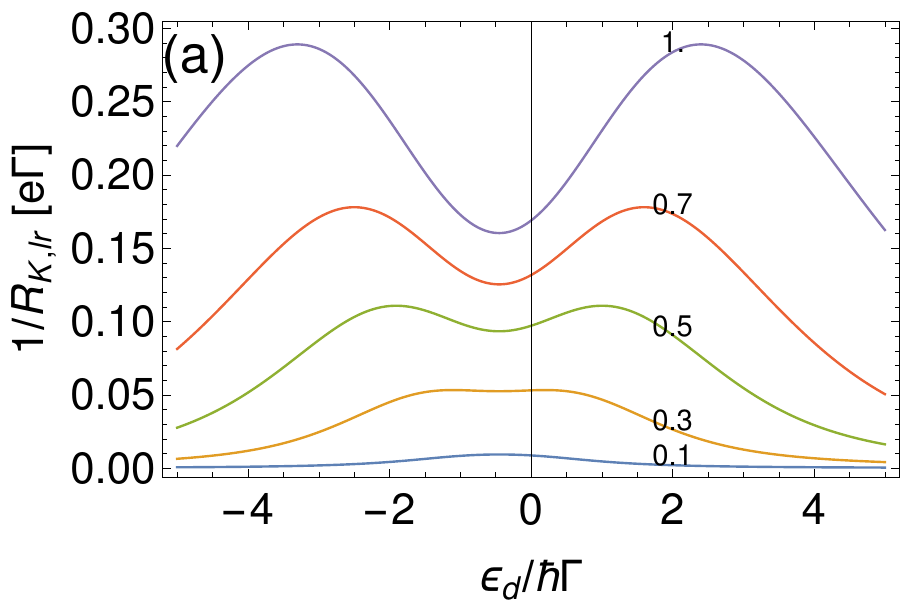}%
  \includegraphics[width=7cm]{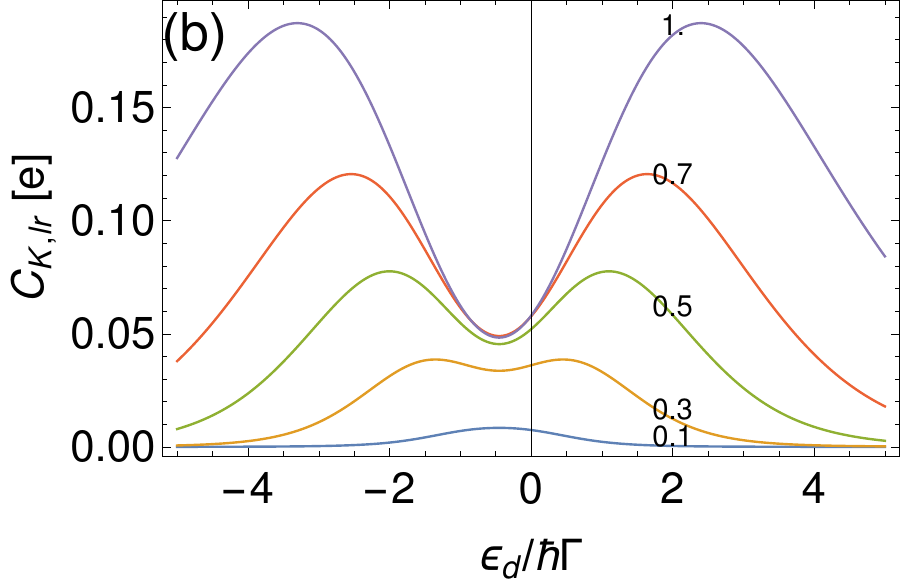}\\
  \includegraphics[width=7cm]{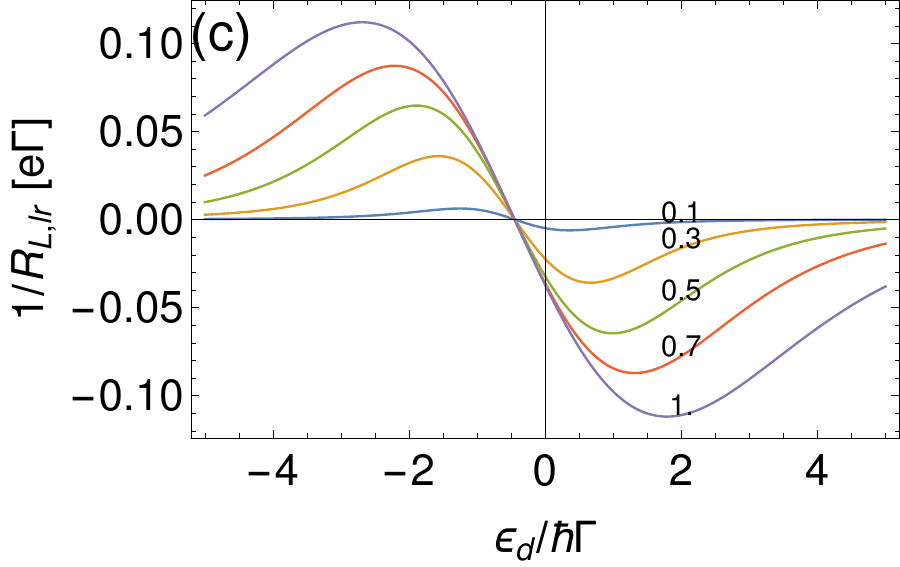}%
  \includegraphics[width=7cm]{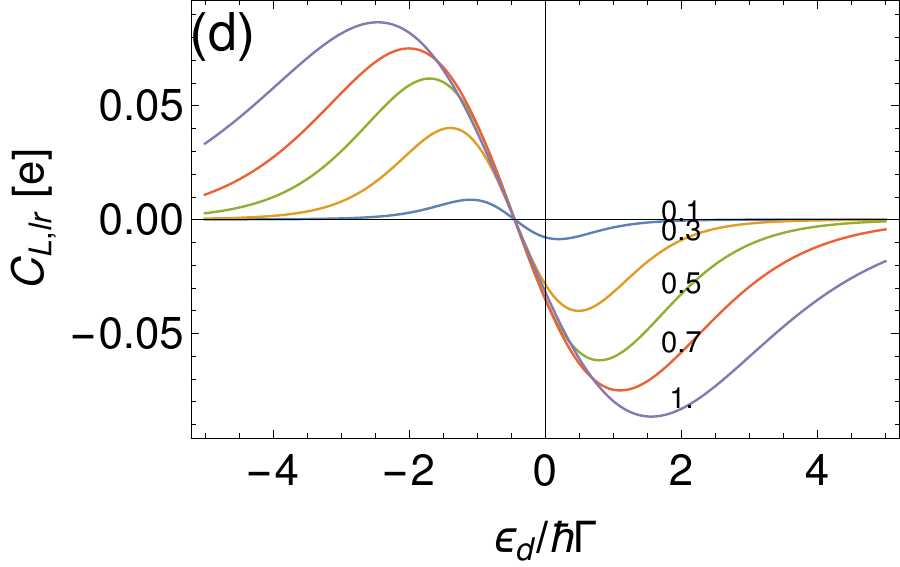}
  \caption{(a) The cross thermal conductance $1/R_{K,\Left\Right}$ and (b) the cross thermal
    capacitance $C_{K,\Left\Right}$, and (c) the cross thermoelectric conductance
    $1/R_{L,\Left\Right}$ and (d) the cross thermoelectric  capacitance $C_{L,\Left\Right}$ in the low-frequency
    limit as functions of the spin-degenerate dot level $\epsilon_d$ in the
    interacting case with $U/\hbar\Gamma = 0.9$ and the symmetric coupling,
    $\Gamma_{\Left} = \Gamma_{\Right}$.  Each curve corresponds to the different
    temperature $k_BT/\hbar\Gamma$ whose value is annotated.}
  \label{fig:ht:LK:cross}
\end{figure*}

Our formalism is not limited to the noninteracting case. The charge and heat
currents, \eqnsref{eq:Ic:linear:final} and (\ref{eq:Ih:linear:final}) can be
calculated as long as the equilibrium and $n=1$ components of the
retarded/advanced QD Green's functions are provided. In this section we take
into account the Coulomb interaction in the quantum dot which is now described
by
\begin{align}
  \label{eq:HD:int}
  \varH_{\rm D}
  =
  \sum_\sigma \epsilon_\sigma d_\sigma^\dag d_\sigma
  + U n_\up n_\down.
\end{align}
Unfortunately, in the presence of finite Coulomb interaction, it is impossible
to obtain any analytical form of the QD NEGFs without a proper
approximation. Here we adopt the simplest approximation: the \textit{Hartree
  approximation} which reduces the two-particle correlations to one-particle
ones as
\begin{align}
  \label{eq:Hartree}
  \Avg{\acomm{d_\sigma(t)}{n_{\bar\sigma}(t') d_\sigma(t')}}
  \approx
  \Avg{n_{\bar\sigma}(t')}
  \Avg{\acomm{d_\sigma(t)}{d_\sigma(t')}}.
\end{align}
Then, the Dyson's equations for the QD NEGFs are found to be basically similar
to those for the noninteracting case but with the retarded/advanced self
energies being now replaced by
\begin{align}
  \label{eq:SigmaRA:Hartree}
  \Sigma^{R/A}_{\sigma,\rm HF}(t,t')
  =
  \Sigma^{R/A}_\sigma(t,t')
  + \delta(t-t') \frac{U}{\hbar} \avg{n_{\bar\sigma}(t)}.
\end{align}
Note that lesser self energy $\Sigma^<_\sigma(t,t')$ remains unchanged compared
to the noninteracting case. This additional term in
$\Sigma^{R/A}_{\sigma,\rm HF}(t,t')$ induces two changes compared to the
noninteracting case: (1) The effective dot level is shifted from the
unperturbed one,
\begin{align}
  \epsilon_\sigma
  \to
  \epsilon_{\sigma,\rm HF}
  = \epsilon_\sigma + \frac{U}{\hbar} n_{\bar\sigma}(0),
\end{align}
where
\begin{align}
  \label{eq:n:eq:Hartree}
  n_\sigma(0)
  =
  \int \frac{d\omega}{2\pi} f(\omega)
  \varG^R_\sigma(0,\omega) (2\Gamma) \varG^A_\sigma(0,\omega)
\end{align}
is the equilibrium QD occupation which should be determined in a
self-consistent way. Note that the equilibrium QD Green's functions now depend
on $\epsilon_{\sigma,\rm HF}$:
\begin{align}
  \label{eq:QDG:eq:Hartree}
  \varG^{R/A}_\sigma(0,\omega)
  = \rcp{\omega - \epsilon_{\sigma,\rm HF}/\hbar \pm i \Gamma}.
\end{align}
(2) The $n=1$ Fourier component of the QD Green's functions are now finite:
\begin{align}
  \label{eq:QDG:1:Hartree}
  \varG^{R/A}_\sigma(1,\omega)
  =
  \varG^{R/A}_\sigma(0,\omega+\Omega)
  \frac{U}{\hbar} n_{\bar\sigma}(1,\Omega)
  \varG^{R/A}_\sigma(0,\omega),
\end{align}
where
\begin{align}
  \label{eq:n:1:Hartree}
  n_\sigma(1,\Omega)
  = \int \frac{d\omega}{2\pi i} \varG^<_\sigma(1,\omega)
\end{align}
is the $n=1$ Fourier component of the QD occupation. By using the charge
conservation, one can obtain the explicit expression of $n_\sigma(1,\Omega)$
which is found to be
\begin{align}
  \label{eq:n:1:Hartree:final}
  n_\sigma(1,\Omega)
  = - \sum_\ell \frac{\Psi_\ell}{2} 2\Gamma_\ell X_\sigma(\Omega)
\end{align}
with
\begin{align}
  \label{eq:X}
  X_\sigma(\Omega)
  \equiv
  \frac{P_{1\sigma}(\Omega)
    + \frac{2\Gamma U}{\hbar} P_{0\sigma}(\Omega) P_{1\bar\sigma}(\Omega)}%
  {1 - \left(\frac{2\Gamma U}{\hbar}\right)^2
    P_{0\sigma}(\Omega) P_{0\bar\sigma}(\Omega)}.
\end{align}
Then, following the recipe proposed in our formalism [see \appref{app:hartree}
for further explanations], the charge and heat currents can be obtained [see
\eqnsref{eq:Ic:Hartree} and (\ref{eq:Ih:Hartree})] and the corresponding
self/cross thermoelectric and thermal admittances are found to be
\begin{subequations}
  \label{eq:LK:ht:self}
  \begin{align}
    \frac{L_\ell(\Omega)}{2i\Gamma_\ell}
    & =
    \Omega e \sum_\sigma
    \left[
      P_{1\sigma}(\Omega)
      +
      \frac{2\Gamma U}{\hbar} P_{0\sigma}(\Omega) X_{\bar\sigma}(\Omega)
    \right],
    \\
    \frac{K_\ell(\Omega)}{2i\Gamma_\ell}
    & =
    \Omega (-\hbar) \sum_\sigma
    \left[
      P_{2\sigma}(\Omega)
      +
      \frac{2\Gamma U}{\hbar} P_{1\sigma}(\Omega) X_{\bar\sigma}(\Omega)
    \right]
  \end{align}
\end{subequations}
and
\begin{subequations}
  \label{eq:LK:ht:cross}
  \begin{align}
    \frac{L_{\Left\Right}(\Omega)}{4 \Gamma_{\Left} \Gamma_{\Right}}
    & =
    e \sum_\sigma
    \left[
      P_{1\sigma}(\Omega)
      +
      \frac{i\Omega U}{\hbar} P_{0\sigma}(\Omega) X_{\bar\sigma}(\Omega)
    \right],
    \\
    \frac{K_{\Left\Right}(\Omega)}{4 \Gamma_{\Left} \Gamma_{\Right}}
    & =
    (-\hbar) \sum_\sigma
    \left[
      P_{2\sigma}(\Omega)
      +
      \frac{i\Omega U}{\hbar} P_{1\sigma}(\Omega) X_{\bar\sigma}(\Omega)
    \right].
  \end{align}
\end{subequations}

\begin{figure*}[t]
  \centering
  \includegraphics[width=7cm]{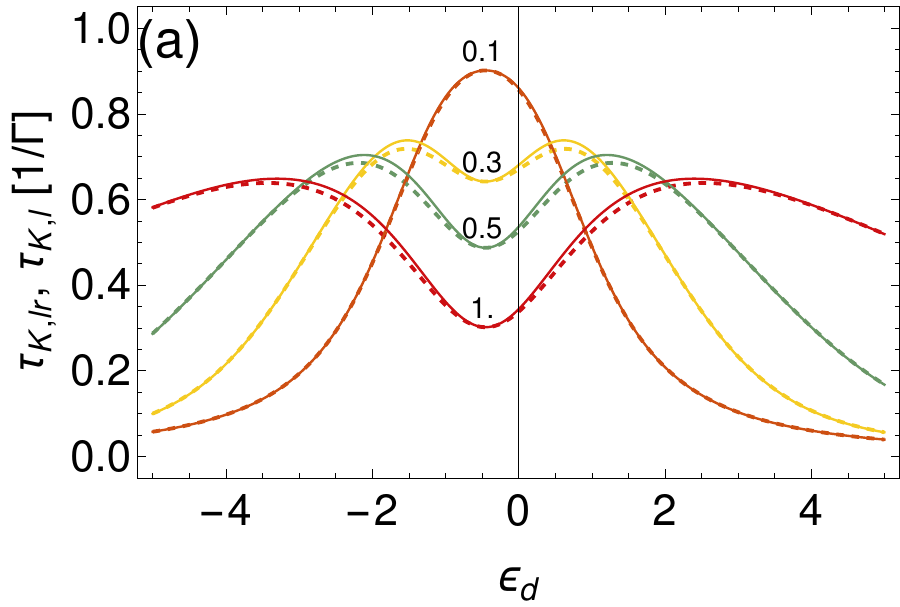}%
  \includegraphics[width=7cm]{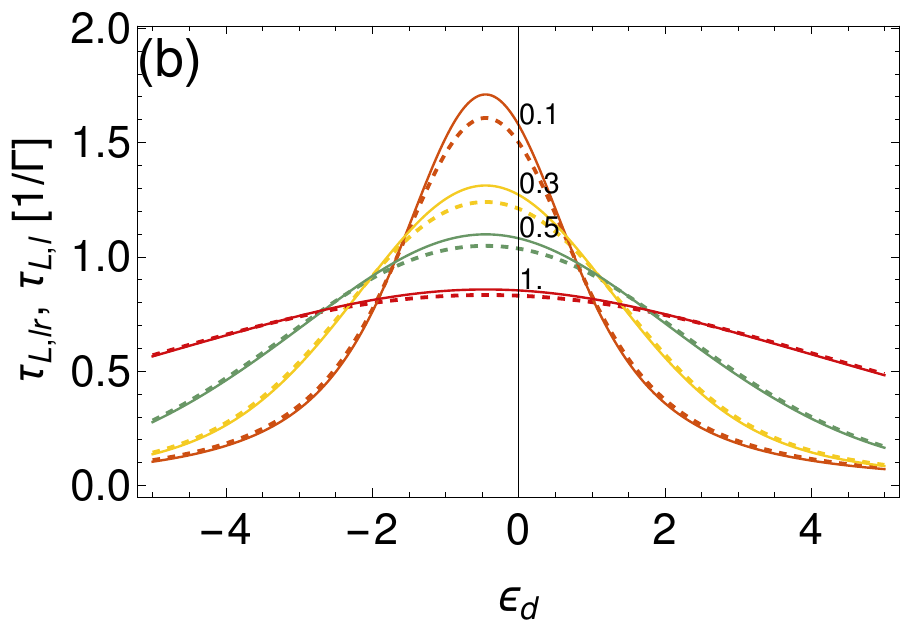}\\
  \includegraphics[width=7cm]{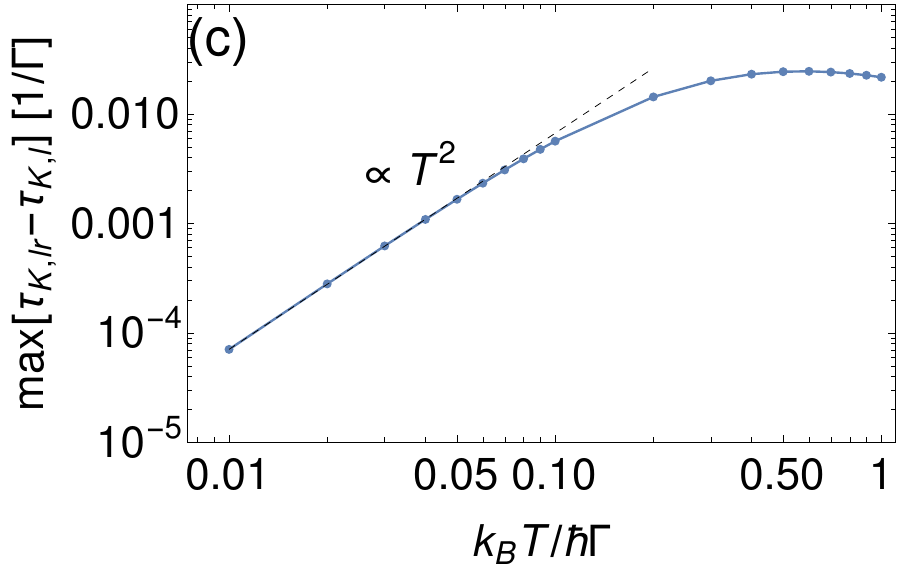}%
  \includegraphics[width=7cm]{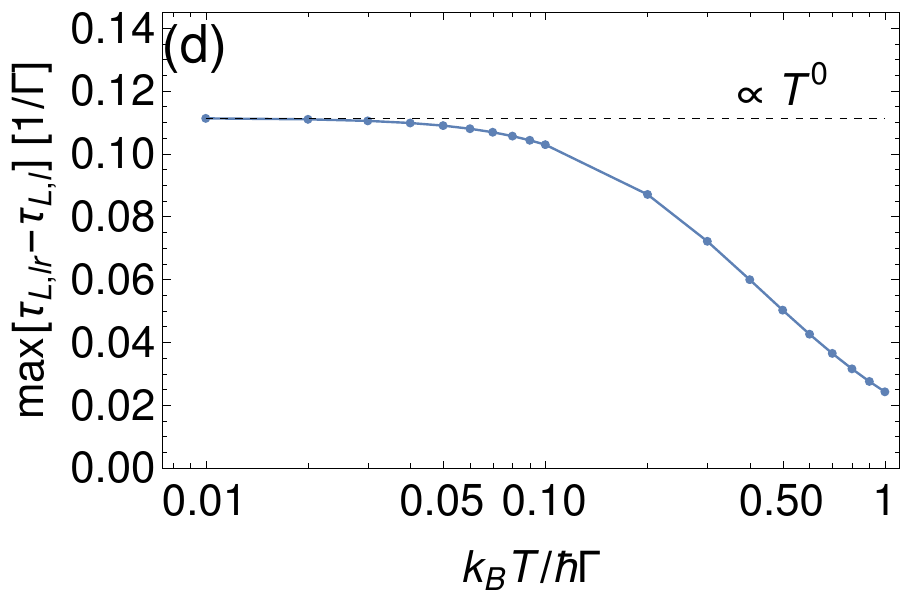}\\
  \caption{(a) The RC times $\tau_{K,\Left\Right}$ (solid lines) and
    $\tau_{K,\Left}$ (dotted lines) from thermal admittance and (b) the RC time
    $\tau_{L,\Left\Right}$ (solid lines) and $\tau_{L,\Left}$ (dotted lines) from
    thermoelectric admittance in the low-frequency limit as functions of the
    spin-degenerate dot level $\epsilon_d$ in the interacting case with the
    same parameters as in \figref{fig:ht:LK:cross}. (c)
    $\max_{\epsilon_d}[\tau_{K,\Left\Right} - \tau_{K,\Left}]$ and (d)
    $\max_{\epsilon_d}[\tau_{L,\Left\Right} - \tau_{L,\Left}]$ as functions of the
    temperature. The dotted lines are asymptotes in the $T\to0$ limit.}
  \label{fig:ht:LK:time}
\end{figure*}

These admittances clearly display the corrections due to the Coulomb
interaction, as shown in \figref{fig:ht:LK:cross}: The resonance is shifted and
the curves are slightly deformed compared to the noninteracting case, but no
qualitative changes are observed. For examples, the low-frequency and
low-temperature cross resistances $R_{L,\Left\Right}$ and $R_{K,\Left\Right}$ are found
to be identical to those in the noninteracting case except $\epsilon_\sigma$
being replaced by $\epsilon_{\sigma,\rm HF}$. Therefore, it is not convenient
to find a solid evidence on the effect of the Coulomb interaction from the
dot-level dependence of the resistances and capacitances.

Instead, we focus on the RC times. In the presence of the Coulomb interaction,
the self and cross RC times are not equal to each other any longer:
\begin{align}
  \tau_{L,\ell} \ne \tau_{L,\Left\Right}
  \quad\text{and}\quad
  \tau_{K,\ell} \ne \tau_{K,\Left\Right},
\end{align}
as demonstrated in \figsref{fig:ht:LK:time}~(a) and (b). The Coulomb
corrections in $Y_\ell$ and $Y_{\Left\Right}$ ($Y = L, K$) are different [compare
\eqnref{eq:LK:ht:self} with \eqnref{eq:LK:ht:cross}] and make their frequency
dependence different from each other. As matter of fact, the difference between
them is proportional to $U$ for weak Coulomb interaction. By performing the
low-frequency and low-temperature expansions similar to those done in the
noninteracting case, one can find that, for spin-degenerate case with
$\rho_\up(0) = \rho_\down(0) \equiv \rho_d$,
\begin{subequations}
  \begin{align}
    \tau_{L,\Left\Right} - \tau_{L,\ell}
    & =
    \frac{hU}{2}
    \left(\frac{\rho_d^2}{1 - U \rho_d} - \frac{\rho_d}{h\Gamma}\right)
    + \varO(T^2)
    \\
    \tau_{K,\Left\Right} - \tau_{K,\ell}
    & =
    \frac{hU}{2} \frac{\pi^2}{3} (k_BT)^2 \rho_d^{\prime2}
    \left(1 - \frac{1}{h\Gamma \rho_d}\right)
    + \varO(T^4).
  \end{align}
\end{subequations}
These expansions show that (1) both $\tau_{L,\Left\Right} - \tau_{L,\ell}$ and
$\tau_{K,\Left\Right} - \tau_{K,\ell}$ are finite in the presence of the Coulomb
interaction and (2) $\tau_{L,\Left\Right} - \tau_{L,\ell}$ saturates in the $T=0$
limit, while $\tau_{K,\Left\Right} - \tau_{K,\ell}$ scales as $T^2$. These asymtotic
behaviors for low temperatures are clearly manifested for a rather wide range
of the temperature, as demonstrated in \figsref{fig:ht:LK:time}~(c) and (d).
We expect that these temperature dependencies of the differences between the RC
times can be used to identify the effect of the Coulomb interaction as long as
$U$ is sufficiently small.

It should be noted that these temperature dependences due to the Coulomb
corrections originate from the dynamical excitations reflected in the $n=1$
components of the QD Green's functions, $\varG^{R/A<}_\sigma(1,\omega)$, and not
from the effective level shift in $\epsilon_{\sigma,\rm HF}$. Our Luttinger
formalism takes into account the dynamical excitations systematically and
correctly, even in the presence of the Coulomb interaction. Therefore, the
utility of our formalism becomes more evident in the interacting systems.

In order to investigate the effect of the strong Coulomb interaction on the
thermoelectric and thermal admittances, one should go beyond the Hartree
approximation, taking into account the higher-order terms in the
equation-of-motion method. For example, as long as the Coulomb blockade is
concerned, one can apply the Meir-Wingreen-Lee approximations
\cite{Meir1991jun,Meir1993apr} to our formalism, which will be our future
work.

\section{Conclusions}
\label{sec:conclusion}

We have formulated a general formalism, based on the Luttinger's trick, to
calculate the dynamic charge and heat currents through a multi-level quantum
dot which is driven by time-dependent temperatures. Our Luttinger formalism is
built on the correct definition of the dynamical contact energy and the
requirement for the effective dot-contact coupling. It provides the general
expressions for the linear-response charge and heat currents, given by
\eqnsref{eq:Ic:linear} and (\ref{eq:Ih:linear:final}), respectively, which can
be calculated as long as the $n=0$ (equilibrium) and $n=1$ Fourier components
of the retarded/advanced and lesser QD Green's functions are
known. Furthermore, with the help of charge conservation and the vanishing sum
of the energy change rates, the knowledge of the lesser QD Green's functions
can be avoided as long as some conditions are met.


The physically important point of our formalism is that it goes beyond the
static and adiabatic condition for temperature modulation and it can capture
naturally and systematically the effect due to dynamical excitations driven by
the nonadiabatic temperature driving. Our formalism is then expected to be very
adequate and useful, considering the state-of-the-art technology which enables
very fast modulation of temperatures.
Furthermore, our formalism works even in the presence of the Coulomb
interaction and can reveal the role of the interaction in nonequilibrium
thermal responses. Its application to the interacting case studied in
\secref{sec:hartree} clearly demonstrates the success of our formalism in this
regard, even though it has been done only in the Hartree approximation.

While the applications of our formalism present in this paper [see
\secsref{sec:ni} and \ref{sec:hartree}] are limited to the single-level
quantum-dot system, it can be applied to any multi-level quantum-dot junctions
so as to study the effect of multiple levels and spin-orbit interactions on the
temperature-driven transport. Since our formalism allows the quantum dot to be
any finite systems, it can also consider the case of multiple quantum dots.
Furthermore, some interferometer-like geometry such as the junction with a
quantum-embedded ring can be also incorporated into our formalism. As long as
one ignores the Coulomb interactions in these systems, the equation-of-motion
method can be readily exploited to derive the required retarded/advanced QD
Green's functions and subsequently the dynamical charge and heat currents.

Real challenge is to take into account the Coulomb interaction in a
non-perturbative way. There is no analytical solution in this case without
proper approximations. Usually, it is very hard to obtain the exact form of
equilibrium QD Green's functions let alone their dynamical ($n=1$) Fourier
components.
One promising method to deal with the Coulomb interaction in a non-perturbative
way is to use the numerical renormalization group \cite{Bulla2008apr} in order
to obtain the QD Green's functions in a numerical way. While originally the
numerical renormalization group is restricted to the equilibrium case, recently
its extension, so-called the time-dependent numerical renormalization group, is
improved to deal with the nonequilibrium case with a periodic driving
\cite{Nghiem2018oct,Nghiem2020mar}, so that the two-time QD Green's functions
are successfully calculated. While it has some issues about the accuracy, we
believe that this method can be safely used to study the linear response regime
in which the driving is sufficiently weak.

Another interesting theoretical challenge is to extend the Luttinger formalism
beyond the linear regime. The original proposal of the Luttinger's trick
\cite{Luttinger1964} was based on the linear response regime so that the
gradient of the gravitational field is identified with the gradient of the
temperature. In fact, there is no physically reliable justification of the use
of the Luttinger's trick beyond the linear regime. However, recently some tried
to extend the Luttinger's idea into the nonlinear-regime study of the
nanostructures: the steady-state and transient behaviors of charge and heat
currents after sudden quench of the gravitational field
\cite{Eich2014,Eich2016} and temperature-driven adiabatic pumping
\cite{Hasegawa2017,Hasegawa2018}. The predictions made by these works are quite
interesting and nontrivial, while their validity of the prediction is uncertain
and to be confirmed in experiments. So, knowing that there is no systematic way
to deal with the time-dependent temperature, it may be very physically
interesting to extend our formalism beyond the linear response regime so that a
new physics is explored.

Finally, we would like to address briefly the experimental realization of our
scheme.  In order to test our predictions, a time-dependent modulation and
control of temperature in the reservoirs should be experimentally
implemented. We think, for example, that our theory may be tested in spin
qubits \cite{Portugal2021nov} with detunnings of order of few meV that
corresponds to ac frequencies for the temperature modulation of about hundreds
of MHz. These are frequencies that are experimentally accessible
nowadays. Alternatively, for higher frequencies (in the GHz and THz range), a
design of time-dependent temperature signals by employing a collection of
quantum harmonic oscillators that mediate the interactions between the quantum
system and a thermal bath has been recently proposed in
Ref.~\cite{Portugal2022nov}.
It is worth noting that in order to realize our prediction, the driving
frequency should be not too low.  It is because our scheme is based on the
quantum-mechanically coherent state during the driving so the period of the
driving should be shorter than the decoherence time, which in turn requires
frequencies in the GHz regime in quantum-dot systems
\cite{Gabelli2006}. In short, the main hurdle to deal with is to maintain the quantum coherence long enough during the
temperature modulation.

\section*{Acknowledgments}

\paragraph{Funding information}
R.L.~acknowledges the financial support by the Grant No. PDR2020/12 sponsored
by the Comunitat Autonoma de les Illes Balears through the Direcció General de
Política Universitaria i Recerca with funds from the Tourist Stay Tax Law ITS
2017-006, the Grant No. PID2020-117347GB-I00, and the Grant No. LINKB20072 from
the CSIC i-link program 2021.  This work has been partially supported by the
María de Maeztu project CEX2021-001164-M funded by the
MCIN/AEI/10.13039/501100011033.
P.S.~acknowledges the financial support of the French National Research Agency
(project SIM-CIRCUIT, ANR-18-CE47-0014-01).
M.L.~was supported by the National Research Foundation of Korea (NRF) grant
funded by the Korea government (MSIT)(No.2018R1A5A6075964).


\begin{appendix}

\section{Nonequilibrium Green functions and Dyson's equations}
\label{app:FG}https://www.overleaf.com/project/63bd2e1bdda5662b33615b26

In our study we employ the nonequilibrium Keldysh formalism to express charge
and heat current in terms of the QD Green's functions. It is convenient to
recast the Green's functions in a matrix form as
\begin{align}
  \widehat\varG
  =
  \begin{bmatrix}
    \varG^t(t,t') & \varG^<(t,t')
    \\
    \varG^>(t,t') & \varG^{\bar{t}}(t,t')
  \end{bmatrix},
\end{align}
where $\varG^<_{\varA,\varB}(t,t') = i\Avg{|\varB^\dag(t') \varA(t)|}$,
$\varG^<_{\varA,\varB}(t,t') = -i\Avg{|\varA(t) \varB^\dag(t')|}$, and \linebreak
$\varG^{t(\bar t)}_{\varA,\varB}(t,t') = -i\Avg{|\varT_{c(\bar c)} \varA(t)
  \varB^\dag(t')|}$ are the lesser, greater, and (anti-)time-ordered Green's
functions between operators $\varA$ and $\varB$, respectively. The retarded and
advanced Green's functions can be obtained through
$\varG^R = \varG^t - \varG^<$ and $\varG^A = \varG^< - \varG^{\bar{t}}$. Here
$\varA$ and $\varB$ can be either the dot operator $d_m$ or the contact
operator $c_{\ell\bfk\sigma}$, and accordingly, the QD, contact, and QD-contact
Green's functions are defined as given in \eqnsref{eq:GF:QD} and
(\ref{eq:GF:contact-QD}). We apply the equation-of-motion technique with
respect to the Hamiltonian (\ref{eq:H}). For convenience, we introduce the
time-varying tunneling amplitudes (\ref{eq:t}) and contact energy
(\ref{eq:epsilon}) so that the contact and tunneling Hamiltonians can be
written as
\begin{subequations}
  \begin{align}
    \varH_{\rm C\ell,\Psi}(t)
    & =
    \sum_{\bfk\sigma} \epsilon_{\ell\bfk}(t)
    c_{\ell\bfk\sigma}^\dag c_{\ell\bfk\sigma},
    \\
    \varH_{\rm T\ell,\Psi}(t)
    & =
    \sum_{m\bfk\sigma}
    \left(
      t_{\ell\bfk\sigma,m}(t) d_m^\dag c_{\ell\bfk\sigma}
      +
      t_{\ell\bfk\sigma,m}^*(t) c_{\ell\bfk\sigma}^\dag d_m
    \right),
  \end{align}
\end{subequations}
apart from the time-dependent numbers which do not affect the dynamics of the
Green's functions.

Via the equation-of-motion method, it is straightforward to obtain the
following Dyson's equations:
\begin{subequations}
  \begin{align}
    \label{eq:Dyson:QDCG}
    \Gm_{m,\ell\bfk\sigma}(t,t')
    & =
    \int dt''\,
    \sum_{m''} \Gm_{m,m''}(t,t'') \frac{t_{\ell\bfk\sigma,m''}(t'')}{\hbar} \tau_3
    \gm_{\ell\bfk\sigma}(t'',t'),
    \\
    \nonumber
    \Gm_{\ell\bfk\sigma,\ell'\bfk'\sigma'}(t,t')
    & =
    \delta_{\ell\ell'} \delta_{\bfk\bfk'} \delta_{\sigma\sigma'} \gm_{\ell\bfk\sigma}(t,t')
    \\
    \label{eq:Dyson:CG}
    & \quad\mbox{}
    +
    \int dt''\, \gm_{\ell\bfk\sigma}(t,t'') \tau_3
    \sum_m \frac{t_{\ell\bfk\sigma,m}^*(t'')}{\hbar}\Gm_{m,\ell'\bfk'\sigma'}(t'',t'),
  \end{align}
\end{subequations}
where $\tau_3$ is the third Pauli matrix in the Keldysh space and
$\gm_{\ell\bfk\sigma}(t,t')$ denotes the \textit{uncoupled} contact Green's
function matrix as introduced in the main text. Explicitly, the uncoupled
contact Green's functions are given by
\begin{subequations}
  \label{eq:gk}
  \begin{align}
    g^{R/A}_{\ell\bfk\sigma}(t,t')
    & =
    \mp i \Theta(\pm(t-t'))
    e^{-\frac{i}{\hbar} \int_{t'}^t dt'' \epsilon_{\ell\bfk}(t'')},
    \\
    g^<_{\ell\bfk\sigma}(t,t')
    & =
    e^{-\frac{i}{\hbar} \int_{t'}^t dt'' \epsilon_{\ell\bfk}(t'')}
    i f(\epsilon_{\ell\bfk}),
  \end{align}
\end{subequations}
where $f(\epsilon)$ is the Fermi function at the temperature $T$ and the phase
factor is evaluated as
\begin{align}
  e^{-\frac{i}{\hbar}\int_{t'}^t dt'' \epsilon_{\ell\bfk}(t'')}
  =
  e^{- \frac{i}{\hbar} \epsilon_{\ell\bfk}
    \left(t - t' + \wPsi_\ell(t) - \wPsi_\ell(t')\right)}
\end{align}
with
\begin{align}
  \wPsi_\ell(t)
  = \int_0^t dt' \Psi_\ell(t')
  = \frac{\Psi_\ell}{\Omega} \sin\Omega t.
\end{align}
In order to evaluate the charge and heat currents, we need to get the
expressions for $\varG^<_{m,\ell\bfk\sigma}(t,t')$ and
$\varG^<_{\ell\bfk\sigma,\ell'\bfk'\sigma'}(t,t')$ [see \eqnsref{eq:Ic:1},
(\ref{eq:HC:1}), and (\ref{eq:HT:1})] from the Dyson's equations
(\ref{eq:Dyson:QDCG}) and (\ref{eq:Dyson:CG}):
\begin{subequations}
  \begin{align}
    \nonumber
    \varG^<_{m,\ell\bfk\sigma}(t,t')
    & =
    \sum_{m''}
    \int dt''\,
    \bigg[
    \varG^R_{mm''}(t,t'') \frac{t_{\ell\bfk\sigma,m''}(t'')}{\hbar}
    g^<_{\ell\bfk\sigma}(t'',t')
    \\
    & \qquad\qquad\qquad\quad\mbox{}
    +
    \varG^R_{mm''}(t,t'') \frac{t_{\ell\bfk\sigma,m''}(t'')}{\hbar}
    g^<_{\ell\bfk\sigma}(t'',t')
    \bigg],
    \\
    \nonumber
    \varG^<_{\ell\bfk\sigma,\ell\bfk\sigma}(t,t')
    & =
    g^<_{\ell\bfk\sigma}(t,t')
    \\
    \nonumber
    & \quad\mbox{}
    +
    \sum_{m''m'''}
    \int dt'' \int dt'''
    \frac{t_{\ell\bfk\sigma,m''}^*(t'')}{\hbar}
    \bigg(
    g^R_{\ell\bfk\sigma}(t,t'')
    \varG^R_{m''m'''}(t'',t''')
    g^<_{\ell\bfk\sigma}(t''',t')
    \\
    \nonumber
    & \qquad\qquad\qquad\mbox{}
    +
    g^<_{\ell\bfk\sigma}(t,t'')
    \varG^A_{m''m'''}(t'',t''')
    g^A_{\ell\bfk\sigma}(t''',t')
    \\
    & \qquad\qquad\qquad\mbox{}
    +
    g^R_{\ell\bfk\sigma}(t,t'')
    \varG^<_{m''m'''}(t'',t''')
    g^A_{\ell\bfk\sigma}(t''',t')
    \bigg)
    \frac{t_{\ell\bfk\sigma,m'''}(t''')}{\hbar}.
  \end{align}
\end{subequations}
We insert the above expressions into \eqnsref{eq:Ic:1}, (\ref{eq:HC:1}), and
(\ref{eq:HT:1}) and obtain \eqnsref{eq:IcHT} and (\ref{eq:HC:2}) in terms of
the relevant self energies $\bfSigma^a_\ell$ defined as \eqnref{eq:Sigma} and
the self-energy-like terms $\bfXi^{ab}_\ell$ defined as \eqnref{eq:Xi}.

\section{Explicit expressions and linear expansions of
  $\bfSigma^{R/A/<}_\ell(t,t')$ and
  $\bfXi^{AR/<R/A<}_\ell(t,t',t'')$}
\label{app:self_energies}

To evaluate the charge and heat currents via \eqnsref{eq:IcHT} and
(\ref{eq:HC:2}), one needs to know the explicit expressions for the self
energies $\bfSigma^a_\ell(t,t')$ [see \eqnref{eq:Sigma}] and the
self-energy-like forms $\bfXi^{ab}_\ell(t,t',t'')$ [see \eqnref{eq:Xi}]. For
simplicity, we take the wide-band limit with a constant density of states
$\rho_0$ for both the contacts, which allows us to replace the sum
$\sum_\bfk F(\epsilon_{\ell\bfk})$ by the integral
$\rho_0\int_{-\infty}^\infty d\epsilon\, F(\epsilon)$.

Using the explicit expressions for $g^{R/A/<}_{\ell\bfk\sigma}(t,t')$, one can
express the self energies $\bfSigma^{R/A/<}_\ell(t,t')$ as
\begin{subequations}
  \label{eq:Sigma:expexp}
  \begin{align}
    \bfSigma^{R/A}_\ell(t,t')
    & =
    \mp 2i\bfGamma_\ell
    \Theta(\pm(t-t'))
    (1+\lambda\Psi_\ell(t))(1+\lambda\Psi_\ell(t'))
    \int \frac{d\omega}{2\pi} e^{-i\omega (t - t' + \wPsi_\ell(t) - \wPsi_\ell(t'))},
    \\
    \bfSigma^<_\ell(t,t')
    & =
    2i\bfGamma_\ell (1 + \lambda\Psi_\ell(t)) (1 + \lambda\Psi_\ell(t'))
    \int \frac{d\omega}{2\pi}
    f(\omega) e^{-i\omega(t-t'+\wPsi_\ell(t)-\wPsi_\ell(t'))},
  \end{align}
\end{subequations}
where $\omega = \epsilon/\hbar$ and we have used $f(\omega)$ instead of
$f(\hbar\omega)$, for simplicity. The integration over $\omega$ can be done for
$\Sigma^{R/A}_{\ell\sigma}$, giving rise to
\begin{align}
  \label{eq:SigmaRA}
  \bfSigma^{R/A}_\ell(t,t')
  =
  \mp i\bfGamma_\ell \frac{(1+\lambda\Psi_\ell(t))^2}{1 + \Psi_\ell(t)}
  \delta(t - t'),
\end{align}
where we have used the fact that $|\Psi_\ell| \le 1$: The temperature
oscillation amplitude cannot be larger than the base temperature $T$
itself. The corresponding Fourier components,
$\bfSigma^{R/A}_{\ell,n}(\omega)$ can be also obtained in an analytical
form (we do not write down the detailed expression here) which involves the
regularized generalized hypergeometric functions. However, unfortunately, no
simple analytical expressions for $\bfSigma^<_\ell(t,t')$ nor
$\bfSigma^<_{\ell,n}(\omega)$ are available. Specifically, by using the
identity
$e^{i\beta\sin\Omega t} = \sum_{n=-\infty}^\infty e^{i\Omega t} J_n(\beta)$
where $J_n(x)$ are the first kind Bessel functions and with a help of
recurrence relations for $J_n(x)$, one can obtain
\begin{align}
  \label{eq:SigmaLesser:exact}
  \bfSigma^<_{\ell,n}(\omega)
  & =
  2i\bfGamma_\ell
  \sum_{m=-\infty}^\infty
  f(\omega_m)
  J_{n+m}(\frac{\Psi_\ell}{\Omega} \omega_m)
  J_m(\frac{\Psi_\ell}{\Omega}\omega_m)
  \left(
    1 + \frac{\lambda(n+m)}{\omega_m} \Omega
  \right)
  \left(
    1 + \frac{\lambda m}{\omega_m} \Omega
  \right)
\end{align}
with $\omega_m \equiv \omega - m\Omega$. For general value of $\Psi_\ell$, this
expression is not adequate for analytical nor numerical analysis since it
requires the summation over $m$ from $-\infty$ to $\infty$.

The situation becomes worse for $\bfXi^{ab}_\ell(t,t',t'')$ whose explicit
expressions involve more complicated integration:
\begin{subequations}
  \label{eq:Xi:expexp}
  \begin{align}
    \nonumber
    \bfXi^{AR}_\ell(t,t',t'')
    & =
    -2i\hbar\bfGamma_\ell
    \Theta(t-t') \Theta(t-t'')
    (1 + \lambda\Psi_\ell(t')) (1 + \lambda\Psi_\ell(t''))
    \\
    & \qquad\mbox{}
    \times
    \int \frac{d\omega}{2\pi} \omega
    e^{-i\omega (t'-t''+\wPsi_\ell(t')-\wPsi_\ell(t''))},
    \\
    \nonumber
    \bfXi^{<R}_\ell(t,t',t'')
    & =
    - 2i\hbar\bfGamma_\ell
    \Theta(t-t'')
    (1 + \lambda\Psi_\ell(t')) (1 + \lambda\Psi_\ell(t''))
    \\
    & \qquad\mbox{}
    \times
    \int \frac{d\omega}{2\pi} \omega f(\omega)
    e^{-i\omega (t'-t''+\wPsi_\ell(t')-\wPsi_\ell(t''))},
    \\
    \bfXi^{A<}_\ell(t,t',t'')
    & = \left[\bfXi_\ell^{<R}(t,t'',t')\right]^\dag.
  \end{align}
\end{subequations}
Unfortunately, for general values of $\Psi_\ell$, the integrations over
$\omega$ and the Fourier transformation do not yield any manageable analytical
expressions. On the other hand, we have found that the linear expansion of the
self energies with respect to $\Psi_\ell$ yields reliable and analytical
expressions. Therefore, in our study we focus on the linear response regime.

Before presenting the linear expansion of the self energies, the reliability of
the linear expansion should be examined. The linear expansion in $\Psi_\ell$
approximates the exponential function in the integrals,
\eqnsref{eq:Sigma:expexp} and (\ref{eq:Xi:expexp}) as
\begin{align}
  \label{eq:linearexp}
  e^{-i\omega(\wPsi_\ell(t) - \wPsi_\ell(t'))}
  \approx
  1 - i\omega(\wPsi_\ell(t) - \wPsi_\ell(t'))
\end{align}
before the integration over $\omega$ is done. While $\Psi_\ell$ is assumed to
be small enough, in fact, $\omega \Psi_\ell$ is not small for large values of
$\omega$ which definitely happen during the integration over $\omega$, which
may disqualify the use of the linear expansion.
However, we have confirmed that this expansion produces correct results. Our
justifications are two-fold.
First, in the linear response regime, only the contact excitations close to the
Fermi level are relevant: Note that $\epsilon = \hbar\omega$ is the contact
excitation energy. Hence, the correctness of the approximation at higher
energies does not matter.
Secondly, we have explicitly adopted a \textit{regularization function}
$\eta(\omega)$ to the gravitational field so that the thermal driving is really
effective only to the low-energy states:
$\Psi_\ell \to \eta(\omega) \Psi_\ell$. Specifically, we have chosen a Gaussian
regularization $\eta(\omega) = \exp[-(\omega/\omega_0)^2]$ with a constant
$\omega_0$ which determines the range of energies to be meaningfully coupled to
the gravitational field. Then, the linear expansion is well justified because
$\omega \eta(\omega) \wPsi(t)$ is small for all values of $\omega$: Note that
$\eta(\omega)$ decreases exponentially with $\omega$. Then, at the final stage
of the calculations, we take $\omega_0\to\infty$ to restore the original
coupling of the gravitational field. We have confirmed that the result obtained
from the regularization and taking the limit of $\omega_0\to\infty$ is
identical to that obtained by using the linear expansion, \eqnref{eq:linearexp}
from the beginning.

Applying the linear expansion (\ref{eq:linearexp}) and performing the Fourier
transformation (\ref{eq:FT}) to $\bfSigma^{R/A/<}_\ell(t,t')$, one can obtain
\begin{align}
  \label{eq:SigmaRA:FT:linear}
  \begin{split}
    \bfSigma^{R/A}_\ell(0,\omega) & = \mp i\bfGamma_\ell,
    \\
    \bfSigma^{R/A}_\ell(1,\omega)
    & = \bfSigma^{R/A}_{\ell\sigma}(-1,\omega)
    = \mp \left(\lambda - \frac12\right) (2i\bfGamma_\ell) \frac{\Psi_\ell}{2}
  \end{split}
\end{align}
and
\begin{align}
  \label{eq:SigmaLesser:FT:linear}
  \begin{split}
    \bfSigma^<_\ell(0,\omega) & = f(\omega) (2i\bfGamma_\ell),
    \\
    \bfSigma^<_\ell(\pm1,\omega)
    & =
    - \Delta_f(\omega\pm\Omega,\omega)
    \left(\omega \pm \frac{\Omega}{2}\right)
    (2i\bfGamma_\ell)
    \frac{\Psi_\ell}{2}
    \\
    & \qquad\quad\mbox{}
    +
    \left(\lambda - \frac12\right) (f(\omega\pm\Omega) + f(\omega))
    (2i\bfGamma_\ell)
    \frac{\Psi_\ell}{2},
  \end{split}
\end{align}
where the definition of $\Delta_f(\omega,\omega')$ is given by
\eqnref{eq:DeltaF}.  As one can see, in the linear response regime, the $n=0$
Fourier components are nothing but the equilibrium values at $\Psi_\ell=0$, and
the $n=\pm1$ components are linear in $\Psi_\ell$ while all the higher
components $(|n|\ge2)$ vanish up to the linear order in $\Psi_\ell$. One can
note that the $n=\pm1$ components become a lot simpler at $\lambda = 1/2$: In
particular, $\bfSigma^{R/A}_\ell(\pm1,\omega)$ vanish at $\lambda = 1/2$.

The linear expansion is now applied to $\Xi^{ab}_\ell(n,n',\omega)$ [see
\eqnref{eq:Xiab:FT}]. Up to the linear order in $\Psi_\ell$, only the Fourier
components with $|n|\le1$, $|n'|\le1$ and $|n+n'|\le1$ are relevant:
\begin{align}
  \label{eq:XiRA:FT:linear}
  \begin{split}
    \bfXi^{AR}_\ell(0,0,\omega)
    & = - I_\infty \hbar\omega (2i\bfGamma_\ell),
    \\
    \bfXi^{AR}_\ell(\pm1,0,\omega)
    & =
    \pm 2(1-\lambda) \frac{i}{\Omega} \hbar\omega
    (2i\bfGamma_\ell)
    \frac{\Psi_\ell}{2},
    \\
    \bfXi^{AR}_\ell(0,\pm1,\omega)
    & =
    \mp
    \frac{i}{\Omega} \hbar \left(\omega \pm \frac{\Omega}{2}\right)
    (2i\bfGamma_\ell)
  \end{split}
\end{align}
and
\begin{align}
  \label{eq:XiRLesser:FT:linear}
  \begin{split}
    \bfXi^{<R}_\ell(0,0,\omega)
    & = - I_\infty f(\omega) \hbar\omega (2i\bfGamma_\ell),
    \\
    \bfXi^{<R}_\ell(\pm1,0,\omega)
    & =
    \pm
    \frac{i}{\Omega}
    \Delta_{\omega f}(\omega\mp\Omega,\omega)
    \hbar \left(\omega \mp \frac{\Omega}{2}\right)
    (2i\bfGamma_\ell)
    \frac{\Psi_\ell}{2}
    \\
    & \quad\mbox{}
    \mp
    \left(\lambda-\frac12\right)
    \frac{i}{\Omega}
    [f(\omega)\hbar\omega + f(\omega\mp\Omega) \hbar(\omega\mp\Omega)]
    (2i\bfGamma_\ell)
    \frac{\Psi_\ell}{2},
    \\
    \bfXi^{<R}_\ell(0,\pm1,\omega)
    & =
    \mp \frac{i}{\Omega}
    f(\omega) \hbar\omega
    (2i\bfGamma_\ell)
  \end{split}
\end{align}
and
\begin{align}
  \label{eq:XiLesserA:FT:linear}
  \begin{split}
    \bfXi^{A<}_\ell(0,0,\omega)
    & = [\bfXi^{<R}_\ell(0,0,\omega)]^\dag,
    \\
    \bfXi^{A<}_\ell(\pm1,0,\omega)
    & =
    [\bfXi^{<R}_\ell(\mp1,0,\omega)]^\dag,
    \\
    \bfXi^{A<}_\ell(0,\pm1,\omega)
    & =
    \left[\bfXi^{<R}_\ell(0,\mp1,\omega\pm\Omega)\right]^\dag
  \end{split}
\end{align}
with $I_\infty \equiv \int_{-\infty}^\infty dt\, \Theta(-t)$. Note that owing
to the constant $I_\infty$ the equilibrium contributions,
$\bfXi^{RA/R</<A}_\ell(0,0,\omega)$ are divergingly large. However, we have
found that these diverging contributions cancel out each other exactly in the
zeroth-order term of the linear expansion of \eqnref{eq:HC:2}:
\begin{align}
  \begin{split}
    (\Avg{\varH_{\rm C\ell}} - E_{\rm C\ell0})^{(0)}
    & =
    \int \frac{d\omega}{2\pi}
    \Tr
    \bigg[
    \bfXi^{AR}_\ell(0,0,\omega) \bfG^<(0,\omega)
    \\
    \nonumber
    & \qquad\qquad\qquad\mbox{}
    +
    \bfXi^{<R}_\ell(0,0,\omega)
    \bfG^R(0,\omega)
    +
    \bfXi^{A<}_\ell(0,0,\omega)
    \bfG^A(0,\omega)
    \bigg]
    \\
    \nonumber
    & =
    I_\infty \int \frac{d\omega}{2\pi} \hbar\omega
    \Tr
    \left[
      (2i\bfGamma_\ell)
      \left(
        \bfG^<(0,\omega)
        +
        f(\omega)
        \left(
          \bfG^R(0,\omega) - \bfG^A(0,\omega)
        \right)
      \right)
    \right]
    \\
    & = 0
  \end{split}
\end{align}
where in the last line we have used the equality,
\begin{align}
  \bfG^<(0,\omega)
  +
  f(\omega)
  \left(
    \bfG^R(0,\omega) - \bfG^A(0,\omega)
  \right)
  = 0
\end{align}
which holds generally \textit{in equilibrium} for any \textit{interacting}
quantum-dot junctions.

\section{Linear regime: charge and heat currents and sum rules}
\label{sec:linear_response_regime:currents}

The charge currents in the linear regime can be obtained by expressing
\eqref{eq:Ic:2} in terms of the Fourier components $\bfG(n,\omega)$ and
$\bfSigma_\ell(n,\omega)$ via \eqref{eq:FTapp:1} and then by keeping terms up
to the linear order in $\Psi_\ell$.  It is then quite straightforward to obtain
the explicit expression, giving rise to \eqnref{eq:Ic:linear}. In the
derivation, we have used the expressions for $\bfSigma^{R/A/<}_\ell(n,\omega)$
[see \eqnsref{eq:SigmaRA:FT:linear} and (\ref{eq:SigmaLesser:FT:linear})] with
$\lambda = 1/2$.

Our system conserves the total charge so that the charge conservation
(\ref{eq:charge_conservation:linear}) holds. As long as the condition
(\ref{eq:simplifying_condition}) holds, the charge conservation condition can
be used to remove $\int d\omega\, \Tr[\bfG^<(1,\omega)]$ from the expression
for the charge current.
By inserting the expressions of the QD charge current (\ref{eq:ID:linear}) and
the contact charge current (\ref{eq:Ic:linear}) into the charge conservation
(\ref{eq:charge_conservation:linear}), one can solve the integral
$\int d\omega \Tr[\bfG^<(1,\omega)]$ in terms of other QD Green's functions:
with $\lambda=1/2$,
\begin{align}
  \label{eq:QDGLesser:linear:cc:final}
  \begin{split}
    & \int \frac{d\omega}{2\pi} \Tr[\bfG^<(1,\omega)]
    \\
    & =
    -
    \int \frac{d\omega}{2\pi}
    \rcp{2i\Gamma + \Omega}
    \bigg[
    -
    \sum_\ell (2i\Gamma_\ell) \frac{\Psi_\ell}{2}
    \Delta_f(\omega+\Omega,\omega) \left(\omega + \frac{\Omega}{2}\right)
    \Tr
    \left[
      \bfG^R(0,\omega+\Omega)
      -
      \bfG^A(0,\omega)
    \right]
    \\
    & \qquad\qquad\qquad\qquad\quad\mbox{}
    +
    f(\omega+\Omega) (2i\Gamma)
    \Tr
    \left[
      \bfG^R(1,\omega+\Omega)
      -
      \bfG^A(1,\omega)
    \right]
    \bigg].
  \end{split}
\end{align}
By inserting \eqnref{eq:QDGLesser:linear:cc:final} into the contact charge
current (\ref{eq:Ic:linear}), one gets \eqnref{eq:Ic:linear:final}.

For obtaining the heat currents we first take the average energies that are
then expanded into
\begin{subequations}
  \begin{align}
    \Avg{\varH_{\rm T\ell}}
    & =
    E_{\rm T\ell0} (1 - \lambda\Psi_\ell(t))
    +
    E_{\rm T\ell}(\Omega) e^{-i\Omega t}
    +
    E_{\rm T\ell}(-\Omega) e^{i\Omega t}
    \\
    \Avg{\varH_{\rm C\ell}}
    & =
    E_{\rm C\ell0}
    +
    E_{\rm C\ell}(\Omega) e^{-i\Omega t}
    +
    E_{\rm C\ell}(-\Omega) e^{i\Omega t}
  \end{align}
\end{subequations}
with
\begin{align}
  \label{eq:ETunnel:linear}
  \begin{split}
    E_{\rm T\ell}(\Omega)
    & =
    \hbar
    \int \frac{d\omega}{2\pi i}
    \Tr
    \bigg[
    \bfSigma^<_\ell(1,\omega)
    \left(
      \bfG^R(0,\omega+\Omega)
      +
      \bfG^A(0,\omega)
    \right)
    \\
    & \qquad\qquad\qquad\quad\mbox{}
    +
    f(\omega+\Omega) (2i\bfGamma_\ell)
    \left(
      \bfG^R(1,\omega+\Omega)
      +
      \bfG^A(1,\omega)
    \right)
    \bigg]
  \end{split}
\end{align}
and
\begin{align}
  \label{eq:EC:linear}
  \begin{split}
    E_{\rm C\ell}(\Omega)
    & =
    \int \frac{d\omega}{2\pi}
    \Tr
    \bigg[
    \bfXi^{<R}_\ell(1,0,\omega+\Omega)
    \left(
      \bfG^R(0,\omega+\Omega) - \bfG^A(0,\omega)
    \right)
    +
    \bfXi^{AR}_\ell(1,0,\omega)
    \bfG^<(0,\omega)
    \\
    & \qquad\qquad\mbox{}
    +
    \bfXi^{<R}_\ell(0,1,\omega+\Omega)
    \left(
      \bfG^R(1,\omega+\Omega) - \bfG^A(1,\omega)
    \right)
    +
    \bfXi^{AR}_\ell(0,1,\omega) \bfG^<(1,\omega)
    \bigg]
  \end{split}
\end{align}
and the tunneling barrier energy at equilibrium
\begin{align}
  E_{\rm T\ell0}
  =
  \hbar
  \int \frac{d\omega}{2\pi i}
  f(\omega)
  \Tr
  \left[
    (2i\bfGamma_\ell)
    \left(
      \bfG^R(0,\omega) + \bfG^A(0,\omega)
    \right)
  \right].
\end{align}
Then, from \eqnsref{eq:Ih:0} and (\ref{eq:W:2}), the contact heat currents and
the energy change rates in the linear regime are expressed in terms of
$E_{\rm C/T\ell}(\Omega)$:
\begin{subequations}
  \begin{align}
    W_{\rm C\ell}(\Omega)
    & = -i\Omega E_{\rm C\ell}(\Omega),
    \\
    W_{\rm T\ell}(\Omega)
    & =
    -i\Omega
    \left(
      E_{\rm T\ell}(\Omega) - \frac{\Psi_\ell}{2} \lambda E_{\rm T\ell0}
    \right),
    \\
    I^h_\ell(\Omega)
    & = W_{\rm C\ell}(\Omega) + \lambda W_{\rm T\ell}(\Omega),
  \end{align}
\end{subequations}
where the Fourier components of the energy change rates in the linear regime
are defined as
\begin{align}
  W_{\rm C/T\ell}(t)
  =
  W_{\rm C/T\ell}(\Omega) e^{-i\Omega t} + W_{\rm C/T\ell}(-\Omega) e^{i\Omega t}.
\end{align}
Employing these expressions and for $\lambda = 1/2$ the Fourier component of
the heat current in the linear regime is obtained as
\eqref{eq:Ih:linear:final}.

We have another similar sum rule for the energy change rates, \eqnref{eq:Wsum},
which is written in the linear regime as \eqnref{eq:Wsum:2}.  While
$W_{\rm D}(t) = d\avg{\varH_{\rm D}}/dt = (i/\hbar)
\Avg{\comm{\varH}{\varH_{\rm D}}}$ requires the specification of
$\varH_{\rm D}$, the other terms in the sum rule can be written as
\begin{align}
  \label{eq:Wsumrule:WW}
  \begin{split}
    & \sum_\ell (W_{\rm C\ell}(\Omega) + W_{\rm T\ell}(\Omega))
    =
    \sum_\ell (-i\Omega)
    \left(
      E_{\rm C\ell}(\Omega)
      + E_{\rm T\ell}(\Omega) - \frac12 E_{\rm T\ell0} \frac{\Psi_\ell}{2}
    \right)
    \\
    & =
    \sum_\ell
    \hbar
    \int \frac{d\omega}{2\pi}
    \Tr
    \bigg[
    (2i\bfGamma_\ell)
    \bigg(
    \frac{\Psi_\ell}{2}
    \Delta_f(\omega+\Omega,\omega)
    \left(\omega + \frac{\Omega}{2}\right)^2
    \left(
      \bfG^R(0,\omega+\Omega) - \bfG^A(0,\omega)
    \right)
    \\
    & \qquad\qquad\qquad\qquad\qquad\quad\mbox{}
    +
    \frac{\Psi_\ell}{2}
    \frac{\Omega}{2}
    \Delta_f(\omega+\Omega,\omega)
    \left(\omega + \frac{\Omega}{2}\right)
    \left(
      \bfG^R(0,\omega+\Omega) + \bfG^A(0,\omega)
    \right)
    \\
    & \qquad\qquad\qquad\mbox{}
    -
    (\omega+\Omega)
    f(\omega)
    \bfG^R(1,\omega)
    +
    \omega
    f(\omega+\Omega)
    \bfG^A(1,\omega)
    -
    \left(\omega+\frac{\Omega}{2}\right)
    \bfG^<(1,\omega)
    \bigg)
    \bigg].
  \end{split}
\end{align}
This expression can be used to write the integral
$\int d\omega\, \omega \Tr[\bfG^<(1,\omega)]$ in terms of other QD Green's
functions, via the sum rule (\ref{eq:Wsum:2}).

Finally, we find the expression for the power of dissipation (\ref{eq:P:0}) in
the linear response regime. Up to the lowest order in $\Psi_\ell$, the power
reads
\begin{align}
  \label{eq:P:linear}
  P(t)
  =
  \sum_\ell \dot\Psi_\ell(t)
  \left[
    e^{-i\Omega t}
    \left(E_{\rm C\ell}(\Omega) + \frac{E_{\rm T\ell}(\Omega)}{2}\right)
    + (c.c.)
  \right]
\end{align}
which is of the second order in $\Psi_\ell$.

\section{Noninteracting Case: QD Green Functions and Charge/Heat Currents}
\label{app:ni}

By applying the equation-of-motion method to the noninteracting Hamiltonian
(\ref{eq:HD:ni}), one can obtain the Dyson's equation for the QD Green's
functions
\begin{align}
  \label{eq:Dyson:QD:ni}
  \Gm_\sigma(t,t')
  =
  \gm_\sigma(t,t')
  +
  \sum_{\ell\bfk} \int dt''
  \Gm_{\sigma,\ell\bfk\sigma}(t,t'')
  \frac{t_{\ell\bfk\sigma,\sigma}^*(t'')}{\hbar}
  \tau_3
  \gm_\sigma(t'',t'),
\end{align}
where $\gm_\sigma(t,t')$ are the unpertubed QD Green's functions whose
explicit expressions are identical to those of the uncoupled contact Green's
functions (\ref{eq:gk}) with the replacement of $\epsilon_{\ell\bfk}(t)$ by
$\epsilon_\sigma$. By combining \eqnsref{eq:Dyson:QD:ni} and
(\ref{eq:Dyson:QDCG}), one can find the Dyson's equation for the QD Green's
functions in terms of the self energies (\ref{eq:Sigma:expexp}),
\begin{align}
  \label{eq:QDG:Dyson}
  \Gm_\sigma(t,t')
  =
  \gm_\sigma(t,t')
  +
  \int dt'' \int dt'''
  \Gm_\sigma(t,t'')
  \tau_3 \widehat\Sigma_\sigma(t'',t''') \tau_3
  \gm_\sigma(t''',t')
\end{align}
or, more specifically, the equations for retarded/advanced QD Green's
functions,
\begin{align}
  \label{eq:QDG:RA}
  \varG^{R/A}_\sigma(t,t')
  = g^{R/A}_\sigma(t,t')
  +
  \int dt'' \int dt'''
  \varG^{R/A}_\sigma(t,t'')
  \Sigma^{R/A}_\sigma(t'',t''')
  g^{R/A}_\sigma(t''',t')
\end{align}
For the linear expansion, it is convenient to express the Dyson's equation in
frequency domain so that
\begin{align}
  \varG^{R/A}_\sigma(t,\omega)
  =
  \left(
    1
    +
    \sum_n e^{in\Omega t}
    \varG^{R/A}_\sigma(t,\omega+n\Omega)
    \Sigma^{R/A}_\sigma(n,\omega)
  \right)
  g^{R/A}_\sigma(\omega).
\end{align}
By using the linear expansions (\ref{eq:SigmaRA:FT:linear}) of
$\Sigma^{R/A}_\sigma(n,\omega)$ and by keeping up to the linear order in
$\Psi_\ell$, the $n=0$ (equilibrium) and $n=1$ components of the
retarded/advanced QD Green's functions are found to be
\begin{subequations}
  \begin{align}
    \varG^{R/A}_\sigma(0,\omega)
    & = \rcp{[g^{R/A}_\sigma(\omega)]^{-1} \pm i \Gamma}
    \\
    \label{eq:QDGRA:1:ni}
    \varG^{R/A}_\sigma(1,\omega)
    & =
    \varG^{R/A}_\sigma(0,\omega+\Omega)
    \Sigma^{R/A}_\sigma(1,\omega)
    \varG^{R/A}_\sigma(0,\omega)
  \end{align}
\end{subequations}
with
$\Sigma^{R/A}_\sigma(1,\omega) \equiv \sum_\ell
\Sigma^{R/A}_{\ell,\sigma}(1,\omega)$.  Note that at $\lambda = 1/2$,
$\Sigma^{R/A}_{\ell,\sigma}(1,\omega)$ is zero up to the linear order in
$\Psi_\ell$ so that $\varG^{R/A}_\sigma(1,\omega)$ also vanishes.
Then, by exploiting the properties of the equilibrium noninteracting QD Green's
functions (\ref{eq:QDGF:ni}) and the vanishingness of
$\Sigma^{R/A}_{\ell,\sigma}(1,\omega)$, one can get the explicit expression for
the charge current from the general formula of the charge current
(\ref{eq:Ic:linear:final}):
\begin{align}
  \label{eq:Ic:ni}
  I^c_\ell(\Omega)
   = \Upsilon_\ell (-e) \sum_\sigma P_{1\sigma}(\Omega)
\end{align}
with
\begin{align}
  \Upsilon_\ell
  \equiv
  - \frac{\Psi_\ell}{2} (2i\Gamma_\ell) \Omega
  +
  \frac{\Psi_\ell - \Psi_{\bar\ell}}{2} (4\Gamma_{\Left} \Gamma_{\Right})
\end{align}
with $\Psi_{\bar{\Left}} = \Psi_{\Right}$ and vice versa.

In order to find the explicit expression for the heat current, one should know
the integrals of $\varG^<_\sigma(1,\omega)$, that is,
$\int d\omega\, \varG^<_\sigma(1,\omega)$ and
$\int d\omega\, \omega \varG^<_\sigma(1,\omega)$. In the noninteracting
case, one can easily derive the linear expansions of the lesser QD Green's
functions $\varG^<_\sigma(t,\omega)$ from the Dyson's equation
(\ref{eq:QDG:Dyson}). However, as proposed in the main text, we instead exploit
the charge conservation (\ref{eq:QDGLesser:linear:cc:final}) and the sum rule
(\ref{eq:Wsum:2}) for the energy change rates in order to express the integrals
of $\varG^<_\sigma(1,\omega)$, that is,
$\int d\omega\, \varG^<_\sigma(1,\omega)$ and
$\int d\omega\, \omega \varG^<_\sigma(1,\omega)$ in terms of the equilibrium
retarded/advanced QD Green's functions. Below we derive their explicit
expressions. From the charge conservation (\ref{eq:QDGLesser:linear:cc:final}),
by using the properties of the equilibrium QD Green's functions
(\ref{eq:QDGF:ni}) and $\varG^{R/A}_\sigma(1,\omega) = 0$, one can get
\begin{align}
  \int \frac{d\omega}{2\pi}
  \varG^<_\sigma(1,\omega)
  = - \sum_\ell \frac{\Psi_\ell}{2} (2i\Gamma_\ell) P_{1\sigma}(\Omega).
\end{align}
Refer the definition of $P_{1\sigma}(\Omega)$ to \eqnref{eq:Pn}. For the
noninteracting QD Hamiltonian, the energy change rate $W_{\rm D}(t)$ is
identified as
\begin{align}
  W_{\rm D}(t)
  = \Ode{t} \sum_\sigma \epsilon_\sigma \avg{n_\sigma(t)}
  =
  \Ode{t} \sum_\sigma \epsilon_\sigma
  \int \frac{d\omega}{2\pi i} \sum_n \varG^<_\sigma(n,\omega) e^{-in\Omega t}
\end{align}
and its Fourier component in the linear regime is found to be
\begin{align}
  \label{eq:WD:ni}
  W_{\rm D}(\Omega)
  =
  - \sum_\sigma \epsilon_\sigma
  \int \frac{d\omega}{2\pi} \Omega \varG^<_\sigma(1,\omega).
\end{align}
Now, by inserting the expressions for $W_{\rm D}(\Omega)$ and the partial sum
(\ref{eq:Wsumrule:WW}) into the sum rule (\ref{eq:Wsum:2}), one can find that
\begin{align}
  \label{eq:intGlesser:ni}
  \int \frac{d\omega}{2\pi}
  \left(\omega + \frac12\right) \varG^<_\sigma(1,\omega)
  = - \sum_\ell \frac{\Psi_\ell}{2} (2i\Gamma_\ell) P_{2\sigma}(\Omega).
\end{align}
By inserting this integral into the general expression for the heat current,
\eqnref{eq:Ih:linear:final}, we obtain the explicit expression for the heat
current for the noninteracting case.
\begin{align}
  \label{eq:Ih:ni}
  I^h_\ell(\Omega)
  = \Upsilon_\ell \hbar \sum_\sigma P_{2\sigma}(\Omega).
\end{align}



\section{Interacting Case: The Hartree Approximation}
\label{app:hartree}

In the presence of the Coulomb interaction described by the interacting
Hamiltonian (\ref{eq:HD:int}), the Dyson's equation for the QD NEGFs has an
additional term proportional to $U$:
\begin{align}
  \label{eq:Dyson:QD:int}
  \begin{split}
    \Gm_\sigma(t,t')
    & =
    \gm_\sigma(t,t')
    +
    \int dt''
    \frac{U}{\hbar}
    \Gm_{d_\sigma,n_{\bar\sigma} d_\sigma}(t,t'')
    \tau_3
    \gm_\sigma(t'',t')
    \\
    & \quad\mbox{}
    +
    \sum_{\ell\bfk} \int dt''
    \Gm_{\sigma,\ell\bfk\sigma}(t,t'')
    \frac{t_{\ell\bfk\sigma,\sigma}^*(t'')}{\hbar}
    \tau_3
    \gm_\sigma(t'',t'),
  \end{split}
\end{align}
where $\Gm_{d_\sigma,n_{\bar\sigma} d_\sigma}(t,t')$ are the Green's functions
between the operators $d_\sigma(t)$ and $n_{\bar\sigma}(t') d_\sigma(t')$. Here
we adopt the Hartree approximation (\ref{eq:Hartree}) so that
\begin{align}
  \varG^{R/A}_{d_\sigma,n_{\bar\sigma} d_\sigma}(t,t')
  \approx
  \varG^{R/A}_\sigma(t,t') \Avg{n_{\bar\sigma}(t')}.
\end{align}
Then, we recover the non-interacting Dyson's equations (\ref{eq:QDG:Dyson}) but
with the self energy being now replaced by the Hartree one defined as
\eqnref{eq:SigmaRA:Hartree}. The Fourier components of the Hartree self
energies in the linear response regime are then
\begin{subequations}
  \begin{align}
    \Sigma^{R/A}_{\sigma,\rm HF}(0,\omega)
    & = \pm i \Gamma + \frac{U}{\hbar} n_{\bar\sigma}(0)
    \\
    \Sigma^{R/A}_{\sigma,\rm HF}(1,\omega)
    & = \frac{U}{\hbar} n_{\bar\sigma}(1,\Omega),
  \end{align}
\end{subequations}
where $n_{\bar\sigma}(0)$ and $n_{\bar\sigma}(1,\Omega)$ are the $n=0$
(equilibrium) and $n=1$ Fourier components of the QD occupation
$\avg{n_\sigma(t)}$, given by \eqnsref{eq:n:eq:Hartree} and
(\ref{eq:n:1:Hartree}), respectively. Hence, the equilibrium retarded/advanced
QD Green's functions are modified accordingly [see \eqnref{eq:QDG:eq:Hartree}]
and, according to \eqnref{eq:QDGRA:1:ni}, the $n=1$ component of the
retarded/advanced QD Green's functions are now finite [see
\eqnref{eq:QDG:1:Hartree}].

Since $\varG^{R/A}_\sigma(\pm1,\omega)$ are now finite, the integrals of
$\int d\omega\, \varG^<_\sigma(1,\omega)$ and
$\int d\omega\, \omega \varG^<_\sigma(1,\omega)$ will have additional
terms. First, from the charge conservation
(\ref{eq:QDGLesser:linear:cc:final}), by using the properties of the
equilibrium QD Green's functions (\ref{eq:QDG:eq:Hartree}) and the explicit
expressions (\ref{eq:QDG:1:Hartree}) for $\varG^{R/A}_\sigma(1,\omega)$, one
can get
\begin{align}
  \int \frac{d\omega}{2\pi}
  \varG^<_\sigma(1,\omega)
  =
  - \sum_\ell \frac{\Psi_\ell}{2} (2i\Gamma_\ell) P_{1\sigma}(\Omega)
  + \frac{U}{\hbar} n_{\bar\sigma}(1,\Omega) (2i\Gamma) P_{0\sigma}(\Omega)
\end{align}
and this integral, combined with \eqnref{eq:QDG:1:Hartree}, can be used to
obtain the explicit expression for $n_\sigma(1,\Omega)$, resulting in
\eqnref{eq:n:1:Hartree:final}.
For the QD Hamiltonian (\ref{eq:HD:int}), the energy change rate $W_{\rm D}(t)$
is identified as
\begin{align}
  W_{\rm D}(t)
  =
  \Ode{t}
  \left(
    \sum_\sigma \epsilon_\sigma \avg{n_\sigma(t)}
    + U \avg{n_\up(t) n_\down(t)}
  \right)
  \approx
  \Ode{t}
  \left(
    \sum_\sigma \epsilon_\sigma \avg{n_\sigma(t)}
    + U \avg{n_\up(t)} \avg{n_\down(t)}
  \right)
\end{align}
in the spirit of the Hartree approximation. Its Fourier component in the linear
regime is then found to be
\begin{align}
  \label{eq:WD:Hartree}
  W_{\rm D}(\Omega)
  =
  - \sum_\sigma \epsilon_{\sigma,\rm HF}
  \int \frac{d\omega}{2\pi} \Omega \varG^<_\sigma(1,\omega).
\end{align}
Now, by inserting the expressions for $W_{\rm D}(\Omega)$ and the partial sum
(\ref{eq:Wsumrule:WW}) into the sum rule (\ref{eq:Wsum:2}), one can find that
\begin{align}
  \label{eq:intGlesser:ni:Hartree}
  \begin{split}
    \int \frac{d\omega}{2\pi}
    \left(\omega + \frac12\right) \varG^<_\sigma(1,\omega)
    & =
    - \sum_\ell \frac{\Psi_\ell}{2} (2i\Gamma_\ell) P_{2\sigma}(\Omega)
    -
    \frac{U}{\hbar} n_{\bar\sigma}(1,\Omega) \Omega P_{1\sigma}(\Omega)
    \\
    & \quad\mbox{}
    -
    \int \frac{d\omega}{2\pi}
    \left(\omega + \frac{\Omega}{2}\right)
    \left(
      f(\omega) \varG^R_\sigma(1,\omega)
      -
      f(\omega+\Omega) \varG^A_\sigma(1,\omega)
    \right).
  \end{split}
\end{align}
By inserting this integral into the general expression for the heat current,
\eqnref{eq:Ih:linear:final}, we can get the explicit expression for the heat
current. Finally, the charge and heat currents in the Hartree approximation are
obtained as
\begin{subequations}
  \begin{align}
    \label{eq:Ic:Hartree}
    I^c_\ell(\Omega)
    & =
    \Upsilon_\ell (-e) \sum_\sigma P_{1\sigma}(\Omega)
    +
    \Upsilon'_\ell
    \frac{U}{\hbar} (-e) \sum_\sigma P_{0\sigma}(\Omega) P_{1\bar\sigma}(\Omega),
    \\
    \label{eq:Ih:Hartree}
    I^h_\ell(\Omega)
    & =
    \Upsilon_\ell \hbar \sum_\sigma P_{2\sigma}(\Omega)
    +
    \Upsilon'_\ell
    \frac{U}{\hbar} \hbar \sum_\sigma P_{1\sigma}(\Omega) P_{1\bar\sigma}(\Omega)
  \end{align}
\end{subequations}
with
\begin{align}
  \Upsilon'_\ell
  \equiv
  - \frac{\Psi_\ell}{2} (2i\Gamma_\ell) \Omega (2\Gamma)
  +
  \frac{\Psi_\ell - \Psi_{\bar\ell}}{2} (4\Gamma_{\Left} \Gamma_{\Right}) (i\Omega).
\end{align}

\end{appendix}


\nolinenumbers


\begin{thebibliography}{10}
\providecommand{\url}[1]{\texttt{#1}}
\providecommand{\urlprefix}{URL }
\expandafter\ifx\csname urlstyle\endcsname\relax
  \providecommand{\doi}[1]{doi:\discretionary{}{}{}#1}\else
  \providecommand{\doi}{doi:\discretionary{}{}{}\begingroup
  \urlstyle{rm}\Url}\fi
\providecommand{\eprint}[2][]{\url{#2}}

\bibitem{Datta2005}
S.~Datta,
\newblock \emph{A New Perspective on Transport (In 2 Parts)},
\newblock Cambridge University Press, New York, 2nd edn.,
\newblock Purdue University, USA (2005).

\bibitem{Wang2007}
L.~Wang and B.~Li,
\newblock \emph{Thermal logic gates: Computation with phonons},
\newblock Physical Review Letters \textbf{99}, 177208 (2007),
\newblock \doi{10.1103/PhysRevLett.99.177208}.

\bibitem{Dhar2008}
A.~Dhar,
\newblock \emph{Heat transport in low-dimensional systems},
\newblock Advances in Physics \textbf{57}, 457 (2008),
\newblock \doi{10.1080/00018730802212077}.

\bibitem{Li2012}
N.~Li, J.~Ren, L.~Wang, G.~Zhang, P.~Hänggi and B.~Li,
\newblock \emph{Colloquium: Phononics: Manipulating heat flow with electronic
  analogs and beyond},
\newblock Reviews of Modern Physics \textbf{84}, 1045 (2012),
\newblock \doi{10.1103/RevModPhys.84.1045}.

\bibitem{BenAbdallah2017}
P.~Ben-Abdallah and S.-A. Biehs,
\newblock \emph{Thermotronics: Towards nanocircuits to manage radiative heat
  flux},
\newblock Zeitschrift für Naturforschung A \textbf{72}(2), 151 (2017),
\newblock \doi{doi:10.1515/zna-2016-0358}.

\bibitem{Benenti2017}
G.~Benenti, G.~Casati, K.~Saito and R.~Whitney,
\newblock \emph{Fundamental aspects of steady-state conversion of heat to work
  at the nanoscale},
\newblock Physics Reports \textbf{694}, 1 (2017),
\newblock \doi{https://doi.org/10.1016/j.physrep.2017.05.008}.

\bibitem{Buttiker1993jun}
M.~B\"uttiker, A.~Pr\^etre and H.~Thomas,
\newblock \emph{Dynamic conductance and the scattering matrix of small
  conductors},
\newblock Physical Review Letters \textbf{70}, 4114 (1993),
\newblock \doi{10.1103/PhysRevLett.70.4114}.

\bibitem{Jauho1994aug}
A.-P. Jauho, N.~S. Wingreen and Y.~Meir,
\newblock \emph{Time-dependent transport in interacting and noninteracting
  resonant-tunneling systems},
\newblock Physical Review B \textbf{50}, 5528 (1994),
\newblock \doi{10.1103/PhysRevB.50.5528}.

\bibitem{Platero2004}
G.~Platero and R.~Aguado,
\newblock \emph{Photon-assisted transport in semiconductor nanostructures},
\newblock Physics Reports \textbf{395}(1), 1 (2004),
\newblock \doi{https://doi.org/10.1016/j.physrep.2004.01.004}.

\bibitem{Kouwenhoven1991sep}
L.~P. Kouwenhoven, A.~T. Johnson, N.~C. van~der Vaart, C.~J. P.~M. Harmans and
  C.~T. Foxon,
\newblock \emph{Quantized current in a quantum-dot turnstile using oscillating
  tunnel barriers},
\newblock Physical Review Letters \textbf{67}, 1626 (1991),
\newblock \doi{10.1103/PhysRevLett.67.1626}.

\bibitem{Howe2021}
H.~Howe, M.~Blumenthal, H.~E. Beere, T.~Mitchell, D.~A. Ritchie and M.~Pepper,
\newblock \emph{Single-electron pump with highly controllable plateaus},
\newblock Applied Physics Letters \textbf{119}, 153102 (2021),
\newblock \doi{10.1063/5.0067428}.

\bibitem{Blumenthal2007}
M.~D. Blumenthal, B.~Kaestner, L.~Li, S.~P. Giblin, X.~J. Janssen, M.~Pepper,
  G.~A. Evans and D.~A. Ritchie,
\newblock \emph{Gigahertz quantized charge pumping},
\newblock Nature Physics \textbf{3}, 343 (2007),
\newblock \doi{10.1038/nphys594}.

\bibitem{Giblin2012}
S.~P. Giblin, M.~Kataoka, C.~H.~W. Barnes, D.~A. Williams, L.~Buckle,
  M.~Pepper, G.~A. Evans and D.~A. Ritchie,
\newblock \emph{Towards a quantum representation of the ampere using single
  electron pumps},
\newblock Nature Communications \textbf{3}, 930 (2012),
\newblock \doi{10.1038/ncomms1916}.

\bibitem{Rossi2014}
A.~Rossi, T.~Tanttu, K.~Y. Tan, I.~Iisakka, R.~Zhao, K.~W. Chan, G.~C.
  Tettamanzi, S.~Rogge, A.~S. Dzurak and M.~Möttönen,
\newblock \emph{An accurate single-electron pump based on a highly tunable
  silicon quantum dot},
\newblock Nano Letters \textbf{14}, 3405 (2014),
\newblock \doi{10.1021/nl500640c}.

\bibitem{Tettamanzi2014}
G.~C. Tettamanzi, R.~Wacquez and S.~Rogge,
\newblock \emph{Charge pumping through a single donor atom},
\newblock New Journal of Physics \textbf{16}, 063036 (2014),
\newblock \doi{10.1088/1367-2630/16/6/063036}.

\bibitem{Grabert1992}
H.~Grabert and M.~Devoret, eds.,
\newblock \emph{Single Charge Tunneling},
\newblock Plenum, New York (1992).

\bibitem{Duprez2021may}
H.~Duprez, F.~Pierre, E.~Sivre, A.~Aassime, F.~D. Parmentier, A.~Cavanna,
  A.~Ouerghi, U.~Gennser, I.~Safi, C.~Mora and A.~Anthore,
\newblock \emph{Dynamical coulomb blockade under a temperature bias},
\newblock Physical Review Research \textbf{3}, 023122 (2021),
\newblock \doi{10.1103/PhysRevResearch.3.023122}.

\bibitem{Blick2005}
R.~H. Blick and M.~Grifoni,
\newblock \emph{Focus on nano-electromechanical systems},
\newblock New Journal of Physics \textbf{7}(1), E06 (2005),
\newblock \doi{10.1088/1367-2630/7/1/E06}.

\bibitem{Gabelli2006}
J.~Gabelli, G.~Fève, J.-M. Berroir, B.~Plaçais, A.~Cavanna, B.~Etienne,
  Y.~Jin and D.~C. Glattli,
\newblock \emph{Violation of kirchhoff's laws for a coherent rc circuit},
\newblock Science \textbf{313}(5786), 499 (2006),
\newblock \doi{10.1126/science.1126940}.

\bibitem{Parmentier2012apr}
F.~D. Parmentier, E.~Bocquillon, J.-M. Berroir, D.~C. Glattli,
  B.~Pla\ifmmode~\mbox{\c{c}}\else \c{c}\fi{}ais, G.~F\`eve, M.~Albert,
  C.~Flindt and M.~B\"uttiker,
\newblock \emph{Current noise spectrum of a single-particle emitter: Theory and
  experiment},
\newblock Physical Review B \textbf{85}, 165438 (2012),
\newblock \doi{10.1103/PhysRevB.85.165438}.

\bibitem{Pekola2013}
J.~Pekola, O.-P. Saira, V.~Maisi, A.~Kemppinen, M.~M{\"o}tt{\"o}nen, Y.~Pashkin
  and D.~Averin,
\newblock \emph{Single-electron current sources: toward a redefined definition
  of the ampere},
\newblock Review of Modern Physics \textbf{85}, 1421 (2013),
\newblock \doi{10.1103/RevModPhys.85.1421}.

\bibitem{Averin2008}
D.~Averin and J.~Pekola,
\newblock \emph{Nonadiabatic charge pumping in a hybrid single-electron
  transistor},
\newblock Physical Review Letters \textbf{101}, 066801 (2008),
\newblock \doi{10.1103/PhysRevLett.101.066801}.

\bibitem{McNeil2007}
R.~McNeil, M.~Kataoka, C.~Ford, C.~Barnes, D.~Anderson, G.~Jones, I.~Farrer and
  D.~Ritchie,
\newblock \emph{On-demand single-electron transfer between distant quantum
  dots},
\newblock Nature \textbf{477}, 439 (2007),
\newblock \doi{10.1038/nature10444}.

\bibitem{Kaestner2008}
B.~Kaestner, V.~Kashcheyevs, S.~Amakawa, M.~Blumenthal, L.~Li, T.~Janssen,
  G.~Hein, K.~Pierz, T.~Weimann, U.~Siegner and H.~Schumacher,
\newblock \emph{Single-parameter nonadiabatic quantized charge pumping},
\newblock Physical Review B \textbf{77}, 153301 (2008),
\newblock \doi{10.1103/PhysRevB.77.153301}.

\bibitem{Feve2007}
G.~F{\`e}ve, A.~Mah{\'e}, J.-M. Berroir, T.~Kontos, B.~Pla{\c{c}}ais, D.~C.
  Glattli, A.~Cavanna, B.~Etienne and Y.~Jin,
\newblock \emph{An on-demand coherent single-electron source},
\newblock Science \textbf{316}(5828), 1169 (2007),
\newblock \doi{10.1126/science.1141243}.

\bibitem{Brunpicard2016}
J.~Brun-Picard, S.~Djordjevic, D.~Leprat, F.~Schopfer and W.~Poirier,
\newblock \emph{Practical quantum realization of the ampere from the elementary
  charge},
\newblock Physical Review X \textbf{6}(4), 041051 (2016),
\newblock \doi{10.1103/PhysRevX.6.041051}.

\bibitem{Gemmer2009}
J.~Gemmer, M.~Michel and G.~Mahler,
\newblock \emph{Quantum Thermodynamics},
\newblock Springer-Verlag Berlin Heidelberg (2009).

\bibitem{Holubec2020}
V.~Holubec, S.~Steffenoni, G.~Falasco and K.~Kroy,
\newblock \emph{Active brownian heat engines},
\newblock Physical Review Research \textbf{2}, 043262 (2020),
\newblock \doi{10.1103/PhysRevResearch.2.043262}.

\bibitem{Martinez2015}
I.~A. Mart{\'i}nez, {\'E}.~Rold{\'a}n, L.~Dinis, D.~Petrov and R.~Rica,
\newblock \emph{Adiabatic processes realized with a trapped brownian particle},
\newblock Physical Review Letters \textbf{114}, 120601 (2015),
\newblock \doi{10.1103/PhysRevLett.114.120601}.

\bibitem{Scopa2019}
S.~Scopa, G.~Landi, A.~Hammoumi and D.~Karevski,
\newblock \emph{Exact solution of time-dependent lindblad equations with closed
  algebras},
\newblock Physical Review A \textbf{99}, 022105 (2019),
\newblock \doi{10.1103/PhysRevA.99.022105}.

\bibitem{Martinez2016}
I.~A. Mart{\'i}nez, {\'E}.~Rold{\'a}n, L.~Dinis, D.~Petrov, J.~Parrondo and
  R.~Rica,
\newblock \emph{Brownian carnot engine},
\newblock Nature Physics \textbf{12}, 67 (2016),
\newblock \doi{10.1038/nphys3518}.

\bibitem{Rossnagel2016}
J.~Ro{\ss}nagel, S.~Dawkins, K.~Tolazzi, O.~Abah, E.~Lutz, F.~Schmidt-Kaler and
  K.~Singer,
\newblock \emph{A single-atom heat engine},
\newblock Science \textbf{352}, 325 (2016),
\newblock \doi{10.1126/science.aad6320}.

\bibitem{Li2004}
B.~Li, L.~Wang and G.~Casati,
\newblock \emph{Thermal diode: Rectification of heat flux},
\newblock Physical Review Letters \textbf{93}, 184301 (2004),
\newblock \doi{10.1103/PhysRevLett.93.184301}.

\bibitem{Otey2010}
C.~Otey, W.~Lau and S.~Fan,
\newblock \emph{Thermal rectification through vacuum},
\newblock Physical Review Letters \textbf{104}, 154301 (2010),
\newblock \doi{10.1103/PhysRevLett.104.154301}.

\bibitem{Li2006}
B.~Li, L.~Wang and G.~Casati,
\newblock \emph{Negative differential thermal resistance and thermal
  transistor},
\newblock Applied Physics Letters \textbf{88}, 143501 (2006),
\newblock \doi{10.1063/1.2191730}.

\bibitem{BenAbdallah2013}
P.~Ben-Abdallah and S.~A. Biehs,
\newblock \emph{Phase-change radiative thermal diode},
\newblock Applied Physics Letters \textbf{103}, 191907 (2013),
\newblock \doi{10.1063/1.4829618}.

\bibitem{BenAbdallah2014}
P.~Ben-Abdallah and S.-A. Biehs,
\newblock \emph{Near-field thermal transistor},
\newblock Physical Review Letters \textbf{112}, 044301 (2014),
\newblock \doi{10.1103/PhysRevLett.112.044301}.

\bibitem{Joulain2015}
K.~Joulain, Y.~Ezzahri, J.~Drevillon and P.~Ben-Abdallah,
\newblock \emph{Modulation and amplification of radiative far field heat
  transfer: Towards a simple radiative thermal transistor},
\newblock Applied Physics Letters \textbf{106}, 133505 (2015),
\newblock \doi{10.1063/1.4916730}.

\bibitem{DiVentra2009}
M.~Di~Ventra, Y.~V. Pershin and L.~O. Chua,
\newblock \emph{Circuit elements with memory: Memristors, memcapacitors, and
  meminductors},
\newblock Proceedings of the IEEE \textbf{97}(10), 1717 (2009),
\newblock \doi{10.1109/JPROC.2009.2021077}.

\bibitem{Portugal2021nov}
P.~Portugal, C.~Flindt and N.~Lo~Gullo,
\newblock \emph{Heat transport in a two-level system driven by a time-dependent
  temperature},
\newblock Phys. Rev. B \textbf{104}, 205420 (2021),
\newblock \doi{10.1103/PhysRevB.104.205420}.

\bibitem{Thouless1983may}
D.~J. Thouless,
\newblock \emph{Quantization of particle transport},
\newblock Physical Review B \textbf{27}, 6083 (1983),
\newblock \doi{10.1103/PhysRevB.27.6083}.

\bibitem{Buttiker1994}
M.~Büttiker, H.~Thomas and A.~Prêtre,
\newblock \emph{Dynamic conductance and the scattering theory of microwave
  resonators},
\newblock Zeitschrift für Physik B: Condensed Matter \textbf{94}(2), 133
  (1994),
\newblock \doi{10.1007/BF01313027}.

\bibitem{Pretre1996sep}
A.~Pr\^etre, H.~Thomas and M.~B\"uttiker,
\newblock \emph{Dynamic admittance of mesoscopic conductors: Discrete-potential
  model},
\newblock Physical Review B \textbf{54}, 8130 (1996),
\newblock \doi{10.1103/PhysRevB.54.8130}.

\bibitem{Brouwer1998oct}
P.~W. Brouwer,
\newblock \emph{Scattering approach to parametric pumping},
\newblock Physical Review B \textbf{58}, R10135 (1998),
\newblock \doi{10.1103/PhysRevB.58.R10135}.

\bibitem{Brandner2020jan}
K.~Brandner and K.~Saito,
\newblock \emph{Thermodynamic geometry of microscopic heat engines},
\newblock Physical Review Letters \textbf{124}, 040602 (2020),
\newblock \doi{10.1103/PhysRevLett.124.040602}.

\bibitem{Bhandari2020oct}
B.~Bhandari, P.~T. Alonso, F.~Taddei, F.~von Oppen, R.~Fazio and L.~Arrachea,
\newblock \emph{Geometric properties of adiabatic quantum thermal machines},
\newblock Physical Review B \textbf{102}, 155407 (2020),
\newblock \doi{10.1103/PhysRevB.102.155407}.

\bibitem{Portugal2022nov}
P.~Portugal, F.~Brange and C.~Flindt,
\newblock \emph{Effective temperature pulses in open quantum systems},
\newblock Physical Review Research \textbf{4}, 043112 (2022),
\newblock \doi{10.1103/PhysRevResearch.4.043112}.

\bibitem{Luttinger1964}
J.~M. Luttinger,
\newblock \emph{Theory of thermal transport coefficients},
\newblock Phys. Rev. \textbf{135}, A1505 (1964),
\newblock \doi{10.1103/PhysRev.135.A1505}.

\bibitem{Hasegawa2018}
M.~Hasegawa and T.~Kato,
\newblock \emph{Effect of interaction on reservoir-parameter-driven adiabatic
  charge pumping via a single-level quantum dot system},
\newblock Journal of the Physical Society of Japan \textbf{87}(4), 044709
  (2018),
\newblock \doi{10.7566/JPSJ.87.044709}.

\bibitem{Eich2014}
F.~G. Eich, A.~Principi, M.~Di~Ventra and G.~Vignale,
\newblock \emph{Luttinger-field approach to thermoelectric transport in
  nanoscale conductors},
\newblock Physical Review B \textbf{90}, 115116 (2014),
\newblock \doi{10.1103/PhysRevB.90.115116}.

\bibitem{Lozej2018aug}
{\v{C}}.~Lozej and T.~Rejec,
\newblock \emph{Time-dependent thermoelectric transport in nanosystems:
  Reflectionless luttinger field approach},
\newblock Physical Review B \textbf{98}, 075427 (2018),
\newblock \doi{10.1103/PhysRevB.98.075427}.

\bibitem{Tatara2015may}
G.~Tatara,
\newblock \emph{Thermal vector potential theory of transport induced by a
  temperature gradient},
\newblock Physical Review Letters \textbf{114}, 196601 (2015),
\newblock \doi{10.1103/PhysRevLett.114.196601}.

\bibitem{Shastry2008}
B.~S. Shastry,
\newblock \emph{Electrothermal transport coefficients at finite frequencies},
\newblock Reports on Progress in Physics \textbf{72}(1), 016501 (2008),
\newblock \doi{10.1088/0034-4885/72/1/016501}.

\bibitem{Eich2016}
F.~G. Eich, M.~Di~Ventra and G.~Vignale,
\newblock \emph{Temperature-driven transient charge and heat currents in
  nanoscale conductors},
\newblock Physical Review B \textbf{93}, 134309 (2016),
\newblock \doi{10.1103/PhysRevB.93.134309}.

\bibitem{Hasegawa2017}
M.~Hasegawa and T.~Kato,
\newblock \emph{Temperature-driven and electrochemical-potential-driven
  adiabatic pumping via a quantum dot},
\newblock Journal of the Physical Society of Japan \textbf{86}(2), 024710
  (2017),
\newblock \doi{10.7566/JPSJ.86.024710}.

\bibitem{Ludovico2014apr}
M.~F. Ludovico, J.~S. Lim, M.~Moskalets, L.~Arrachea and D.~S\'anchez,
\newblock \emph{Dynamical energy transfer in ac-driven quantum systems},
\newblock Physical Review B \textbf{89}, 161306 (2014),
\newblock \doi{10.1103/PhysRevB.89.161306}.

\bibitem{Ludovico2016}
M.~F. Ludovico, L.~Arrachea, M.~Moskalets and D.~Sánchez,
\newblock \emph{Periodic energy transport and entropy production in quantum
  electronics},
\newblock Entropy \textbf{18}(11) (2016),
\newblock \doi{10.3390/e18110419}.

\bibitem{note1}
A potential argument could state that the appropriate selection of the contact
  energy operator is $\varQ_\ell = \varH_{\rm C\ell,\Psi}(t) + \lambda
  \varH_{\rm T\ell,\Psi}(t)$, since it incorporates the sources, that is, the
  contribution to the energy from the coupling terms of the gravitional fields.
  Evaluating this alongside \eqnref{eq:Q:0}, we observe their difference being
  second order in $\Psi_\ell$. As we are operating in a linear response regime,
  both definitions offer equivalent outcomes for the heat current to this
  order. The determination of the correct contact energy operator definition
  requires a study into the nonlinear regime, which currently surpasses our
  research scope. Besides, the power definition is of second order in
  $\Psi_\ell$, which means that the source contribution to the heat fluxes is
  of second order and it is consistent with the indistinguishability between
  the two candidate definitions for the contact energy.

\bibitem{Langreth1976}
D.~C. Langreth,
\newblock \emph{Linear and nonlinear electron transport in solids},
\newblock In J.~T. Devreese and V.~E. Van~Doren, eds., \emph{Nato Advanced
  Study Institute, Series 8: Physics}, vol.~17. Plenum, New York (1976).

\bibitem{Tien1963Jan}
P.~K. Tien and J.~P. Gordon,
\newblock \emph{Multiphoton process observed in the interaction of microwave
  fields with the tunneling between superconductor films},
\newblock Phys. Rev. \textbf{129}, 647 (1963),
\newblock \doi{10.1103/PhysRev.129.647}.

\bibitem{Rossello2015}
G.~Rossell\'o, R.~L\'opez and J.~S. Lim,
\newblock \emph{Time-dependent heat flow in interacting quantum conductors},
\newblock Physical Review B \textbf{92}, 115402 (2015),
\newblock \doi{10.1103/PhysRevB.92.115402}.

\bibitem{Onsager1931}
L.~Onsager,
\newblock \emph{Reciprocal relations in irreversible processes. i.},
\newblock Phys. Rev. \textbf{37}, 405 (1931),
\newblock \doi{10.1103/PhysRev.37.405}.

\bibitem{Lim2013nov}
J.~S. Lim, R.~L{\'{o}}pez and D.~S{\'{a}}nchez,
\newblock \emph{{Dynamic thermoelectric and heat transport in mesoscopic
  capacitors}},
\newblock Physical Review B \textbf{88}(20), 201304 (2013),
\newblock \doi{10.1103/PhysRevB.88.201304}.

\bibitem{Averin2010}
D.~V. Averin and J.~P. Pekola,
\newblock \emph{Violation of the fluctuation-dissipation theorem in
  time-dependent mesoscopic heat transport},
\newblock Physical Review Letters \textbf{104}, 220601 (2010),
\newblock \doi{10.1103/PhysRevLett.104.220601}.

\bibitem{Sergi2011}
D.~Sergi,
\newblock \emph{Energy transport and fluctuations in small conductors},
\newblock Physical Review B \textbf{83}, 033401 (2011),
\newblock \doi{10.1103/PhysRevB.83.033401}.

\bibitem{Meir1991jun}
Y.~Meir, N.~Wingreen and P.~Lee,
\newblock \emph{Transport through a strongly interacting electron system:
  Theory of periodic conductance oscillations},
\newblock Physical Review Letters \textbf{66}(23), 3048 (1991),
\newblock \doi{10.1103/PhysRevLett.66.3048}.

\bibitem{Meir1993apr}
Y.~Meir, N.~Wingreen and P.~Lee,
\newblock \emph{{Low-temperature transport through a quantum dot: The Anderson
  model out of equilibrium}},
\newblock Physical Review Letters \textbf{70}(17), 2601 (1993),
\newblock \doi{10.1103/PhysRevLett.70.2601}.

\bibitem{Bulla2008apr}
R.~Bulla, T.~Costi and T.~Pruschke,
\newblock \emph{{Numerical renormalization group method for quantum impurity
  systems}},
\newblock Reviews of Modern Physics \textbf{80}(2), 395 (2008),
\newblock \doi{10.1103/RevModPhys.80.395}.

\bibitem{Nghiem2018oct}
H.~T.~M. Nghiem and T.~A. Costi,
\newblock \emph{{Time-dependent numerical renormalization group method for
  multiple quenches: Towards exact results for the long-time limit of
  thermodynamic observables and spectral functions}},
\newblock Physical Review B \textbf{98}(15), 155107 (2018),
\newblock \doi{10.1103/PhysRevB.98.155107}.

\bibitem{Nghiem2020mar}
H.~T.~M. Nghiem, H.~T. Dang and T.~A. Costi,
\newblock \emph{{Time-dependent spectral functions of the Anderson impurity
  model in response to a quench with application to time-resolved photoemission
  spectroscopy}},
\newblock Physical Review B \textbf{101}(11), 115117 (2020),
\newblock \doi{10.1103/PhysRevB.101.115117}.

\end{thebibliography}
\end{document}